\documentclass[a4paper,11pt,hyper]{JHEP3}
\usepackage{amsfonts,latexsym,graphicx,epsfig,amssymb,amsmath,mathrsfs}

\newcommand{\bm}[1]{\hbox{\boldmath{$#1$}}}
\newcommand{\sbm}[1]{\hbox{\boldmath{\scriptsize$#1$}}}

\newcommand{\dd}{{\rm d}}
\newcommand{\sR}{{^s\!R}}
\newcommand{\Seff}{S_{\rm eff}}
\newcommand{\gbz}{{^g\!\bar{\zeta}}}

\newcommand{\Seffd}[1]{S'_{{\rm eff}(#1)}}
\newcommand{\Scf}{S_{\chi,0}}

\newcommand{\gauge}{{\em gauge} }

\title{Conservation of $\zeta$ with radiative corrections from heavy field}
\author{
Takahiro Tanaka$^{a,b}$, Yuko Urakawa$^{c,d}$\\
a.~Department of Physics, Kyoto university,
 Kyoto, 606-8502, Japan\\
b.~Yukawa Institute for Theoretical Physics, Kyoto university,
  Kyoto, 606-8502, Japan
c.~Department of Physics and Astrophysics, Nagoya University, Nagoya
464-8602, Japan\\
d.~School of Natural Sciences, Institute for Advanced Study, Olden Lane, 
Princeton, NJ 08540, USA
}

\abstract{In this paper, we address a possible impact of radiative
corrections from a heavy scalar field $\chi$ on the curvature
perturbation $\zeta$. Integrating out $\chi$, we derive the effective
action for $\zeta$, 
which includes the loop corrections of the heavy field $\chi$. When the
mass of $\chi$ is much larger than the Hubble scale $H$, the loop
corrections of $\chi$ only yield a local contribution to the effective
action and hence the effective action simply gives an action for $\zeta$
in a single field model, where, as is widely known, $\zeta$ is conserved
in time after the Hubble crossing time. Meanwhile, 
when the mass of $\chi$ is comparable to $H$, the loop corrections of
$\chi$ can give a non-local contribution to the effective
action. Because of the non-local contribution from $\chi$, in general,
$\zeta$ may not be conserved, even if the classical background
trajectory is determined only by the evolution of the inflaton. In this
paper, we derive the condition that $\zeta$ is conserved in time in the
presence of the radiative corrections from $\chi$. Namely, we show that
when the scaling symmetry, which is a part of the diffeomorphism
invariance, is preserved at the quantum level, the loop
corrections of the massive field $\chi$ do not disturb the constant evolution of $\zeta$
at super Hubble scales. In this discussion, we show the Ward-Takahashi
identity for the scaling symmetry, which yields a consistency relation
for the correlation functions of the massive field $\chi$. }

\keywords{Inflation, Effective field theory, Adiabatic modes, Consistency relation}
\preprint{}

\maketitle

\begin{document}


\section{Introduction}
Inflation provides us with a natural experimental instrument to explore
the high energy physics. Measurements of the temperature anisotropies
and polarization of the cosmic microwave background can constrain the
Hubble parameter $H$ at the time when the fluctuation was generated. The
current data puts an upper bound on $H$ at around $10^{14}$ GeV~\cite{Ade:2015lrj, Ade:2015tva}, which is
much higher than the accessible energy scale in particle
accelerators. The precise measurements of the primordial perturbations
generated during inflation may place a constraint 
on the theory of high energy physics independently of the particle experiments.

In string theory, compactification of the extra dimensions typically
yields a number of scalar fields, which may have masses bigger than the
Hubble parameter during inflation. Investigating a possible imprint of these massive fields
might allow us to explore the high energy physics behind. While one
field model is consistent with the current data~\cite{Ade:2015lrj},
there is still room to include a contamination of such massive fields,
which act as isocurvature modes. If such a massive field has a mass much bigger than the Hubble scale,
integrating out the massive field only gives local contributions to the
effective action for the inflaton (relevant works can be found, e.g., in
Refs.~\cite{Tolley:2009fg, Jackson:2010cw}). In such a case, since we are ignorant
of the high energy theory, it is impossible to disentangle the
radiative corrections of the massive field. However, if one of the
isocurvature modes has a mass of order $H$, the radiative correction may yield
a distinctive non-local contribution.

Chen and Wang studied an impact of a massive field on the primordial
curvature perturbation $\zeta$ in Ref.~\cite{CW09} (see also Ref.~\cite{Green:2013rd}). In their setup,
the inflaton has a non-minimal coupling with the massive field, which yields the
cross-correlation between them. As emphasized in Ref.~\cite{NJ15}, where
a more extensive analysis, including higher spin fields, was done, the
massive field leaves more direct information in the squeezed
configuration of the correlation functions, which has a soft external leg, than in other
configurations. The massive scalar field with $0< m/H \leq 3/2$ decays
as $\eta^{\Delta_-}$ with
\begin{align}
 & \Delta \equiv \frac{3}{2} - \sqrt{\frac{9}{4}- \frac{m^2}{H^2}}
 \label{Delta-} 
\end{align}
at large scales. Then, as was computed in Refs.~\cite{Green:2013rd, NJ15,
Noumi:2012vr} the contribution of the massive field to the squeezed bi-spectrum, when the shorter mode $k$ crosses the Hubble scale, is given by
\begin{align}
 & \langle \zeta_q \zeta_k \zeta_k \rangle \propto P_\zeta(q) P_\zeta(k)
(q/k)^{\Delta}\,, \qquad \quad (q/k \ll 1)\,,  \label{Eq:bizeta}
\end{align}
where $P_\zeta(k)$ is the power spectrum of $\zeta$. Notice that 
$(q/k)^{\Delta}$ encodes the evolution between the Hubble crossing
time for the mode $k$ and the one for $q$. For $m > 3H/2$, the massive field oscillates, while
decaying, as $\eta^{\tilde\Delta_{\pm}}$ with
\begin{align}
 & \tilde{\Delta}_{\pm} \equiv \frac{3}{2} \pm i \sqrt{\frac{m^2}{H^2}-
 \frac{9}{4}}\,, \label{Deltapm}
\end{align}
which gives the same momentum dependence as in
Eq.~(\ref{Eq:bizeta}) except that $\Delta_-$ is replaced with
$\tilde{\Delta}_\pm$.

When the curvature perturbation stops evolving after the Hubble crossing
time of the shorter mode $k$, Eq.~(\ref{Eq:bizeta}) gives the squeezed bispectrum at the
end of inflation. As far as the massive fields do
not contribute to the background evolution (an example where a massive
field modulates the background evolution was studied, e.g., in
Refs.~\cite{Achucarro:2010jv, Saito12, Saito13, Renaux-Petel:2015mga}) and the tree level
contribution is concerned, the curvature perturbation is conserved after all
modes cross the Hubble scale~\cite{WeinbergAd, WMLL, MW03, LMS, LV05,
NS}\footnote{The conservation of
$\zeta$ in such a setup also can be understood as a direct consequence of
the $\delta N$ formalism~\cite{Starobinsky:1986fxa, Salopek:1990jq, SS,
Sasaki:1998ug}.}. In this case, Eq.~(\ref{Eq:bizeta}) indeed gives the bi-spectrum
at the end of inflation~\cite{CW09, NJ15, MM15}. The argument in Ref.~\cite{NJ15} is based on the symmetry of the de
Sitter spacetime. Therefore, one may speculate that the loop correction
of the massive field may also be still given by Eq.~(\ref{Eq:bizeta}), while the scaling
dimension $\Delta$ will no longer be given as in Eq.~(\ref{Delta-}) nor
Eq.~(\ref{Deltapm}). In this generalization, a non-trivial point may be in
showing the conservation of $\zeta$ after the Hubble crossing. In
Refs.~\cite{SZ1210, ABG}, the conservation of $\zeta$ was addressed, including the
loop correction of $\zeta$ in the setup of single field inflation.

In this paper, we address the conservation of the curvature
perturbation $\zeta$ which is affected by loop corrections of a heavy
field $\chi$, assuming that the heavy field does not contribute to the
classical background trajectory. The constant non-decaying mode
of $\zeta$ is called the adiabatic mode. To compute the evolution of the
curvature perturbation, we integrate out the heavy field and
derive the effective action for $\zeta$. If the mass of the heavy
field $M$ is much bigger than the Hubble scale, the loop corrections of
$\chi$ only yields local terms in the effective action and then
following the argument in the single field case, we can show the
conservation of $\zeta$ at large scales. On the other hand, if the mass
$M$ is not large enough compared to the Hubble scale, the loop
corrections of $\chi$ can give non-local contributions to the effective
action. The presence of the non-local contribution can yield a
qualitative difference from single field models.

In single field models of inflation, the conservation of the curvature
perturbation at large scales is implemented by the scaling symmetry
$\bm{x} \to e^s \bm{x}$ with a constant parameter $s$, which changes
$\zeta(t, \bm{x})$ to $\zeta(t,\, e^{-s}\bm{x})-s$. The scaling symmetry is one of the
gauge transformations and hence classically it should be preserved for a
diffeomorphism invariant theory. However, when we quantize the system,
the scaling symmetry is not always preserved, particularly when we allow
an arbitrary initial quantum state~\cite{IRgauge_L, IRgauge}. When the
scaling symmetry is preserved, a part of the IR divergences is canceled
out~\cite{IRgauge_L, IRgauge, BGHNT10, GHT11, GS10, GS11, SZ1203, PSZ, IRsingle, SRV1,
SRV2}. (In order to eliminate all the IR divergences, we also need to 
preserve the invariance under other gauge
transformations. For a detailed explanation, see, e.g.,
Ref.~\cite{IRreview}.) In Refs.~\cite{SRV2, SRVGW}, it was shown that
when we choose the Euclidean vacuum, a.k.a., the adiabatic vacuum or the 
Bunch-Davies vacuum in the de Sitter limit, there exists a set of
quantities which is free from the IR divergences.

In quantum field theory, a symmetry implies a corresponding identity,
the so-called Ward-Takahashi (WT) identity. For one
field model of inflation, the scaling symmetry yields the consistency
relation, which relates the $(n+1)$-point function of $\zeta$ with one
soft external leg to the $n$-point function of
$\zeta$~\cite{Maldacena02, PM04, HHK}. The consistency relation is
indeed the WT identity for the scaling symmetry~\cite{HHK}. The consistency relation was
first shown for the bi-spectrum in the squeezed limit by Maldacena in
Ref.~\cite{Maldacena02} and it was extended to more general single field
models in Ref.~\cite{PM04}. The consistency relation for the arbitrary
$n$-point function was derived in 
Ref.~\cite{HHK}. In a single field inflation with diffeomorphism
invariance, when the initial state is the Euclidean vacuum and the
background trajectory is on attractor, the consistency relation
generally holds. When one of these assumptions is not fulfilled, the
consistency relation can be violated~\cite{NFS, CNFS, Flauger:2013hra}.

Various extensions of the consistency relation have been attempted so
far. In Refs.~\cite{PM04, GK, RP}, the squeezed bi-spectrum was computed in a 
non slow-roll setup and in Refs.~\cite{Creminelli:2011rh, PJM},
sub-leading contributions for the consistency relation were computed. 
(See also Refs.~\cite{ABD12, SZ12, Joyce:2014aqa, Berezhiani:2013ewa}.)
The consistency relation can also be obtained from the reparametrization
invariance of the wave function of the universe~\cite{Pimentel:2013gza,
Ghosh:2014kba, Kundu:2015xta}. The use of the wave function is also
motivated by the holographic description of inflation~\cite{Maldacena02,
Kundu:2015xta, Mata:2012bx, Garriga:2013rpa, Garriga:2014ema, Garriga:2014fda}. In
Refs.~\cite{Shiu, Bzowski:2012ih}, the consistency relation was derived
by solving the Callan-Symanzik equation in the dual boundary theory (see
also Ref.~\cite{McFadden:2014nta}). A possible gauge issue for the
consistency relation was discussed in Refs.~\cite{IRNG, Creminelli:2011sq, Pajer:2013ana}.

In this paper, we derive the consistency relation for the heavy field
$\chi$ from the requirement of the scaling symmetry. When the scaling symmetry is preserved
at the quantum level, we obtain
the corresponding WT identity. The WT identity for the scaling
symmetry yields the consistency relation which relates the $(n+1)$-point
function of the $n$ $\chi$s and one soft curvature perturbation
$\zeta$ to the $n$-point function of the $\chi$ field. The
derivation of the consistency relation also applies in the
presence of the loop corrections of the heavy field. Using the effective action for
$\zeta$, obtained by integrating out $\chi$, we show that when the
consistency relation for $\chi$ holds, the curvature perturbation
$\zeta$ is conserved at the super Hubble scales.

This paper is organized as follows. In Sec.~\ref{Sec:single}, we review
the conservation of $\zeta$ in single field models of inflation,
emphasizing the crucial role of the scaling symmetry. In Sec.~\ref{Sec:EA}, after we describe our setup of the problem,
we introduce the effective action for $\zeta$ by integrating out the
heavy field in the in-in (or closed time path)
formalism. In Sec.~\ref{Sec:WT}, we derive the consistency relation for
the heavy field from the Ward-Takahashi identity for the scaling
symmetry. In Sec.~\ref{Sec:Conservation}, using the consistency
relation, derived in Sec.~\ref{Sec:WT}, we show the conservation of
$\zeta$ in the presence of the loop corrections of the heavy field. In
Sec.~\ref{Renormalization}, we briefly discuss the renormalization of
the heavy field. Finally, in Sec.~\ref{Conclusion}, we conclude.

\section{Conservation of $\zeta$ and scaling symmetry in single field inflation} \label{Sec:single}
In single field models of inflation, it is known that the curvature
perturbation is conserved in the large scale limit. In this section, we
show that the conservation of $\zeta$ is a direct consequence of the
scaling symmetry. 

\subsection{Single field inflation}  \label{SSec:single}
For illustrative purpose, we start our discussion by considering a single
scalar field with the standard kinetic term, whose action is given by
\begin{align}
 S &= \frac{1}{2} \int \sqrt{-g} \left[ R - g^{\mu \nu}
 \partial_\mu \phi \partial_\nu \phi  - 2V(\phi) \right] \dd^4 x\,.  \label{Exp:actionS}
\end{align}
In this paper, we set the gravitational constant $\kappa^2 \equiv 8\pi
G$ to $1$. Using the ADM form of the line element: 
\begin{align}
 \dd s^2 = - N^2 \dd t^2  + h_{ij} (\dd x^i + N^i \dd t) (\dd x^j + N^j \dd
  t)~,
 \label{Exp:ADMmetric}
\end{align}
where we introduced the lapse function $N$, the
shift vector $N^i$, and the spatial metric $h_{ij}$, we can express the
action (\ref{Exp:actionS}) as  
\begin{align}
 S &=\frac{1}{2} \int \sqrt{h} \Bigl[ N\,\sR - 2 N
  V(\phi) + N (\kappa_{ij} \kappa^{ij} - \kappa^2) \cr
  & \qquad \qquad \qquad \qquad + \frac{1}{N} ( \dot{\phi}
  - N^i \partial_i \phi )^2 - N h^{ij} \partial_i \phi \partial_j \phi
  \Bigr] \dd^4x~, \label{Eq:ADMaction}
\end{align}
where $\sR$ is the three-dimensional Ricci scalar, and $\kappa_{ij}$ and
$\kappa$ are the extrinsic curvature and its trace, defined by 
\begin{align}
 \kappa_{ij} = \frac{1}{2 N} ( \dot{h}_{ij} - D_i N_j
 - D_j N_i )~, \qquad  
 \kappa = h^{ij} \kappa_{ij} ~.
\end{align}
Here, the spatial indices $i, j, \cdots$ are raised or lowered by the spatial
metric $h_{ij}$, and $D_i$ denotes the covariant differentiation 
associated with $h_{ij}$. Taking the variation of the action with
respect to $N$ and $N^i$, which are the Lagrange multipliers, we obtain
the Hamiltonian and momentum constraint equations as
\begin{align}
 & \sR - 2 V -  (\kappa^{ij} \kappa_{ij} - \kappa^2 ) 
          -  N^{-2} ( \dot{\phi} - N^i \partial_i \phi)^2 - h^{ij}
 \partial_i \phi \partial_j \phi = 0~,   \\
 & D_j ( \kappa^j_{~i} - \delta^j_{~i} \kappa ) - N^{-1} 
  \partial_i \phi~ ( \dot{\phi} - N^j \partial_j \phi) = 0~. 
\end{align}

We determine the time slicing, employing the uniform field gauge:
\begin{align}
 & \delta \phi=0\,.   \label{GCphi}
\end{align} 
We express the spatial metric $h_{ij}$ as
\begin{align}
 & h_{ij}= e^{2(\rho+\zeta)} \left[ e^{\delta \gamma} \right]_{ij}\,,
\end{align}
where the background scale factor is expressed as $a \equiv e^{\rho}$ and 
$\delta \gamma_{ij}$ is set to traceless. As spatial gauge conditions,
we impose 
\begin{align}
 & \partial^i \delta \gamma_{ij} =0\,. \label{GCT}
\end{align}
With this gauge choice, the constraint equations are given by 
\begin{align}
 & \sR - 2 V -  (\kappa^{ij} \kappa_{ij} - \kappa^2 ) 
 - N^{-2}  \dot{\phi}^2 = 0\,,   \\ 
 & D_j ( \kappa^j_{~i} - \delta^j_{~i} \kappa ) = 0 ~.
\end{align}
Inserting $N$ and $N_i$, which are expressed in terms of $\zeta$ by
solving these constraint equations, into the action (\ref{Eq:ADMaction}), we can derive the
action for $\zeta$~\cite{Maldacena02, SeeryLidsey05}.

\subsection{Scaling symmetry}  \label{SSec:scaling}
The transverse condition imposed on $\delta \gamma_{ij}$ is non-local
and hence to determine the coordinates, we need to employ boundary
conditions. For example, at linear order in perturbation, the tensor
perturbation transforms under the spatial coordinate transformation 
$x^i \to \delta \tilde{x}^i =x^i + \delta x^i$ as
\begin{align}
 & \delta \tilde{\gamma}_{ij} (x) = \delta \gamma_{ij}(x) - 2
 \left(\partial_{(i} \delta x_{j)} - \frac{1}{3} \delta_{ij} \partial^k
 \delta x_k \right) \,. 
\end{align}  
The transverse condition on $\delta \gamma_{ij}$ gives
\begin{align}
 & \partial^2 \delta x^i = - \frac{1}{3} \partial^i \partial_j \delta
 x^j\,, \label{Exp:condT}
\end{align}
which does not determine $\delta x^i$ uniquely without specifying boundary
conditions to solve Eq.~(\ref{Exp:condT}). For the scalar mode, all
spatial coordinate transformations $\delta x^i=\partial^i\delta x$
which satisfy $\partial^2 \partial^i \delta x=0$ still keep the
transverse condition after the transformations. When we consider a
compact support on each time slicing, we find an infinite way to
impose the boundary conditions in solving 
$\partial^2 \partial^i \delta x=0$. We analyzed these \gauge degrees of
freedom in detail in Refs.~\cite{IRgauge_L, IRgauge}, where we
used the italic font for ``gauge'' to discriminate the \gauge
transformations defined within the compact support from those in the
infinite spatial support. (See Ref.~\cite{HHK} for a more recent work.)  
So far, we kept the tensor perturbation $\delta \gamma_{ij}$ for the
illustrative purpose, but in the rest of this paper, we neglect it.

Among these transformations, the important one for $\zeta$ is the
scale transformation $\delta x^i= s x^i$, which is extended to
$x^i \to x_s^i= e^s x^i$ at non-linear orders. The parameter
$s$ can vary in time~\cite{IRgauge_L, IRgauge}, but for our purpose, we
do not need to consider the time dependence, and hence we set $s$ to a
constant parameter. Under this transformation, the spatial line element
is recast into
\begin{align}
 & \frac{\dd l^2_d}{a^2(t)} = e^{2 \zeta(t,\, \sbm{x})} \dd\bm{x}^2 = e^{2 
 \zeta_s(t,\, \sbm{x}_s)} \dd \bm{x}_s^2 = e^{2 
 \{ \zeta_s(t,\, e^s \sbm{x}) +s \}} \dd \bm{x}^2 \,,
\end{align}
and then the curvature perturbation changes to 
\begin{align}
 & \zeta_s(t,\, \bm{x}) = \zeta(t,\, e^{-s}\bm{x}) -s\,.  \label{Exp:transzeta}
\end{align}
This is a purely geometrical argument, and hence this transformation law
also should apply to multi-field models.

Preserving the scaling symmetry is crucial to ensure the infrared (IR)
regularity of the curvature perturbation. In
Refs.~\cite{IRgauge_L, IRgauge}, we introduced the spatial Ricci scalar
evaluated in the geodesic normal coordinates as a \gauge invariant
quantity. Since the contribution from the IR modes, which give rise to
the IR divergence, can be eliminated by performing the corresponding \gauge
transformation, the IR divergence also can be removed from the \gauge
invariant quantity.

By construction, the spatial Ricci scalar evaluated in the spatial geodesic
normal coordinates is \gauge invariant and it serves a conceptually
clear example of the \gauge invariant quantity. Nevertheless, using the geodesic
normal coordinates can alter the UV behaviour~\cite{Tsamis:1989yu,
Miao:2012xc}. For a practical use, we may use the smeared geodesic
coordinates $\bm{x}_g(t)$ given by~\cite{SRV1, SRV2, SRVGW} 
\begin{align}
 & \bm{x}_g(t) \equiv e^{\gbz(t)} \bm{x} \,, \label{Def:xg}
\end{align} 
where $\gbz$ is the averaged $\zeta$ at a compact support on each time slicing, given by
\begin{align}
 & \gbz(t) = \frac{\int \dd^3 \bm{x}_g W_t(\bm{x}_g) \zeta(t,\,
 e^{- \gbz} \bm{x}_g)}{\int \dd^3 \bm{x}_g W_t(\bm{x}_g)} \,, \label{Def:gbz}
\end{align}
where $W_t(\bm{x})$ is a window function which vanishes outside the
compact support on the time constant slicing. The spatial Ricci scalar evaluated at $\bm{x}_g$ is not
invariant under all the \gauge transformations, but it is invariant
under the scale transformation.

\subsection{Conservation of $\zeta$ in single field inflation}
In single field inflation, solving the Hamiltonian and momentum
constraint equations, we can eliminate $N$ and $N_i$ and write down the
action only in terms of $\zeta$. Since the action 
for any diffeomorphism invariant theory remains invariant under the
scale transformation, the action for $\zeta$ should take the following form:
\begin{align}
 & S [\zeta] = \int \dd t\, \dd^3 \bm{x} e^{3(\rho+\zeta)} {\cal
 L}_\zeta\! \left[ \partial_t\zeta,\, e^{-(\rho+\zeta)} \partial_i \zeta
 \right]\,,  \label{Exp:actionzeta}
\end{align}
where the Lagrangian density ${\cal L}_\zeta$ includes $\zeta$ only in
the form $\partial_t \zeta$ or $e^{-(\rho+\zeta)}\partial_i \zeta$. (A
detailed explanation can be found in Ref.~\cite{SRV1}.)

To address the conservation of $\zeta$ in the large scale limit, we
neglect the terms which include $\zeta$ with the spatial
derivative. Then, the action for $\zeta$, written in the form
(\ref{Exp:actionzeta}), is given by 
\begin{align}
 & S [\zeta]  \approx \int \dd t\, \dd^3 \bm{x} e^{3(\rho+\zeta)}
 \varepsilon  \left[ \dot{\zeta}^2 + \sum_{n=3}^\infty \frac{2}{n} f_n(t)
 \dot{\zeta}^n \right]\,,   \label{Exp:actionlargescale}
\end{align} 
where we schematically wrote the non-linear terms which include
$\dot{\zeta}$. Here, the time dependent function $f_n(t)$ is expressed 
only in terms of the background quantities such as $\dot{\rho}$ and the
slow-roll parameters. Varying the action with respect to $\zeta$, we
obtain the equation of motion in the large scale limit as
\begin{align}
 & \partial_t \left[ e^{3(\rho+\zeta)} \varepsilon \left\{ \dot{\zeta} +
  \sum_{n=3}^\infty  f_n(t) \dot{\zeta}^{n-1} \right\} \right] \approx 0 \,.
 \label{Exp:eomlargescale} 
\end{align}
This equation motion has the anticipated constant solution in time as the
non-decaying mode. I.e., if $\zeta(x)=F(x)$ is a solution 
of Eq.~(\ref{Exp:eomlargescale}), $F(x)+ C$ with a constant
shift should also satisfy the equation. Then, after other solutions of
the non-linear equation decay, $\zeta$ should be conserved at large scales. The
relation between the shift symmetry and the conservation of $\zeta$ was
pointed out also in Horndeski's theory~\cite{Gao:2011mz}.

\section{Effective action for $\zeta$ with loop corrections of heavy
 field}  \label{Sec:EA}
Next, we consider a two-field model with one inflaton and one heavy
field. The latter field does not contribute to the classical background
evolution. Following the Feynman and Vernon's method~\cite{FV}, in this 
section, we compute the effective action for the curvature perturbation
with loop corrections of the massive field.

\subsection{Two field model}  \label{SSec:setup}
In this paper, we consider a light scalar field $\phi$ and a massive
scalar field $\chi$ whose action is given by  
\begin{align}
 S &= \frac{1}{2} \int \sqrt{-g} \left[ R - g^{\mu \nu}
 \partial_\mu \phi \partial_\nu \phi  - g^{\mu \nu}
 \partial_\mu \chi \partial_\nu \chi - 2V(\phi,\, \chi) \right] \dd^{d+1} x\,,  \label{Exp:action}
\end{align}
where $V(\phi,\, \chi)$ is a potential for the scalar fields:
\begin{align}
 & V(\phi,\, \chi) = V_{ph}(\phi) + V_{ch}(\phi,\, \chi)   \,, \label{Exp:V}
\end{align}
with
\begin{align}
 & V_{ch}(\phi,\, \chi) \equiv  \frac{1}{2} M^2(\phi) \chi^2 +
 \frac{\lambda}{4!}\chi^4 \,.
\end{align}
We decomposed the potential $V(\phi,\, \chi)$ into the $\chi$
independent part $V_{ph}$ and the rest $V_{ch}$. 
When the mass of $\chi$ field, $M$, depends on the inflaton
$\phi$, this model includes the direct interaction between $\phi$ and
$\chi$, e.g., $\phi^2\chi^2$, which was addressed in
Refs.~\cite{Wu:2006xp, KKT}. We assume that the mass of the inflaton $m$ is much smaller
than the Hubble parameter $H$ as $m \ll H$, while the mass of the field
$\chi$ is bigger than $H$ as $M > H$. Then, the classical background
evolution is determined solely by the inflaton $\phi$ and (the linear
perturbation of) $\chi$ becomes the pure isocurvature perturbation. In this paper, we only allow
the renormalizable interactions for $\chi$, while $W(\phi)$ may
include non-renormalizable interactions. To perform the dimensional
regularization, we consider a general $(d+1)$ dimensional spacetime. 
An extension to include non-renormalizable interactions for $\chi$ is
straightforward as long as a finite order of loop corrections is concerned.

As in the single field case, we determine the time slicing, imposing 
\begin{align}
 & \delta \phi=0\,.
\end{align} 
For our later use, we discriminate the part which explicitly depends on $\chi$ from
the rest as
\begin{align}
  S= S_{ad} [N,\, N_i,\, \zeta] + S_\chi[N,\, N_i,\, \zeta,\, \chi]\,,
\end{align}
where $S_{ad}$ agrees with the action in single field models, given in
Eq.~(\ref{Eq:ADMaction}), and $S_\chi$ is given by 
\begin{align}
  S_\chi  & =  \frac{1}{2} \int e^{d(\rho+ \zeta)}
 \biggl[ \frac{1}{N} (\dot{\chi} - e^{-2(\rho +\zeta)} \delta^{ij} N_i
 \partial_j \chi)^2 \cr
 & \qquad \qquad \qquad \qquad \quad - N e^{-2(\rho+
 \zeta)} (\partial_i \chi)^2 - 2 N V(\phi,\, \chi) \biggr] \dd t\, \dd^d \bm{x}\,.
\end{align}
Even if there is no direct interaction between $\phi$ and $\chi$, 
gravity yields the non-linear interaction between $\zeta$ and the
heavy field $\chi$.

\subsection{Effective action}
Since the mass of the field $\chi$ is much bigger than the Hubble scale
$H$, it is natural to set the background value of $\chi$ to 0. Then, 
the field $\chi$ does not contribute to the classical background
evolution. Meanwhile, because of the interaction between $\zeta$ and 
$\chi$, the quantum fluctuation of $\chi$ can affect the evolution of
the field $\zeta$. In order to compute the evolution of $\zeta$ under
the influence of $\chi$, we compute the Feynman and Vernon's influence
functional~\cite{FV, FH, CL1982}, which can be obtained by integrating out the
field $\chi$, in the closed time path (or the in-in) formalism. 

As in the single field case, the action $S$ also includes the lapse
function and the shift vector, which can be removed by solving the
Hamiltonian and momentum constraint equations. These constraint
equations are also modified due to the quantum fluctuation of the
heavy field $\chi$. 

Using the closed time path, the $n$-point function of the curvature
perturbation is given by 
\begin{align}
 & \langle \Psi\,| T\zeta(x_1) \cdots \zeta(x_n) |\, \Psi \rangle\cr
 & = \frac{\int D\zeta_+ \int D\chi_+ \int D \zeta_- \int D\chi_-\,
 \zeta_+(x_1) \cdots \zeta_+(x_n)\, e^{i S[\delta g_+,\, \chi_+]-
 i S[\delta g_-,\, \chi_-]}}{\int D\zeta_+ \int D\chi_+ \int D \zeta_- \int
 D\chi_-\, e^{i S[\delta g_+,\, \chi_+]-
 i S[\delta g_-,\, \chi_-]}}\,,  
\end{align}
where we used an abbreviation $\delta g= (\delta N,\, N_i,\, \zeta)$. In
the closed time path, we double the fields: $\delta g_+$ and
$\chi_+$ denote the fields integrated from the past infinity to the time $t$ and $\delta g_-$ and $\chi_-$ denote the
fields integrated from the time $t$ to the past infinity. Here, $T$
denotes the time ordering. Inserting $\zeta_-(x)$ into the path integral
in the numerator, we can compute the $n$-point function ordered in the
anti-time ordering. Since $N$ and $N_i$ are not independent variables, we
perform the path integral only regarding $\zeta$ and $\chi$.

Introducing the effective action $\Seff$ as
\begin{align}
 &i \Seff[\zeta_+,\, \zeta_-] \equiv  \ln \left[ \int D \chi_+ \int D \chi_- \,
 e^{i S[\delta g_+,\, \chi_+]- i S[\delta g_-,\, \chi_-]}\right]\,, \label{Def:Seff}
\end{align}
we rewrite the $n$-point function for $\zeta$ as
\begin{align}
 &  \langle \Psi\,| T\zeta(x_1) \cdots \zeta(x_n) |\, \Psi \rangle
 = \frac{\int D\zeta_+  \int D \zeta_- \, \zeta_+(x_1) \cdots
 \zeta_+(x_n)\, e^{i \Seff[\delta g_+,\,
 \delta g_-]}}{\int D\zeta_+ \int D \zeta_-\,e^{i
 \Seff[\delta g_+,\,\delta g_-]}} \,.   
\end{align}
By inserting the action $S$ into Eq.~(\ref{Def:Seff}),
the effective action is recast into
\begin{align}
 & \Seff[\delta g_+,\, \delta g_-] = S_{ad}[\delta g_+] -
 S_{ad}[\delta g_-] + \Seff'[\delta g_+,\, \delta g_-]\,, \label{Exp:Seff}
\end{align}
where $\Seff'$ is the so-called influence functional, given by 
\begin{align}
 &i   \Seff'[\delta g_+,\, \delta g_-] 
 \equiv \ln \left[ \int D \chi_+ \int D \chi_- \,e^{i S_\chi[\delta g_+,\,
 \chi_+]- i S_\chi[\delta g_-,\, \chi_-]}\right]\,,  \label{Exp:Seffd}
\end{align}
where we factorized $S_{ad}[\delta g_\pm]$ which commute with the path integral
over $\chi_\pm$. The effective action $\Seff[\delta g_+,\, \delta g_-]$ describes the
evolution of the curvature perturbation affected by the quantum
fluctuation of the heavy field $\chi$.

\subsection{Computing the effective action}  \label{SSec:NaiveEA}
Performing the path integrals about $\chi_+$ and $\chi_-$, we can
compute the effective action $\Seff'[\delta g_+,\, \delta g_-]$. Expanding
$\Seff'$ in terms of $\delta g= (\delta N,\, N_i,\, \zeta)$, we obtain 
\begin{align}
 & i \Seff'[\delta g_+,\, \delta g_-] \equiv \sum_{n=0}^\infty i
 \Seffd{n}[\delta g_+,\, \delta g_-] \,,
\end{align}
where $\Seffd{n}$ denotes the terms which include $n$
$\delta g_\alpha$s, given by
\begin{align}
 i \Seffd{n}[\delta g_+,\, \delta g_-] 
 &= \frac{1}{n!}
 \sum_{\alpha_1 = \pm} \cdots \sum_{\alpha_n = \pm}
 \int \dd^{d+1} x_1 \cdots \int \dd^{d+1} x_n \cr
 & \qquad \times \delta g_{\alpha_1}(x_1) \cdots
 \delta g_{\alpha_n}(x_n)\,W^{(n)}_{\delta g_{\alpha_1}
 \cdots \delta g_{\alpha_n}}(x_1,\, \cdots,\, x_n) \,, \label{Expn:Seffd}
\end{align} 
with
\begin{align}
 & W^{(n)}_{\delta g_{\alpha_1} \cdots \delta g_{\alpha_n}}(x_1,\, \cdots,\, x_n) \equiv
 \frac{\delta^n i \Seff'[\zeta_+,\, \zeta_-]}{\delta g_{\alpha_1}(x_1)
 \cdots \delta  g_{\alpha_n}(x_n)}
 \bigg|_{\delta g_{\pm}=0} \,.  \label{Def:tWn}
\end{align}
In Eq.~(\ref{Expn:Seffd}), each $\delta g_{\alpha_m}$ with 
$m=1,\cdots,\, n$ should add up $\delta N_{\alpha_m}$, $N_{i, \alpha_m}$, and $\zeta_{\alpha_m}$. Here and
hereafter, for notational brevity, we omit the summation symbol over
$\delta g$ unless necessary. Using Eq.~(\ref{Exp:Seffd}), we can express the variation of $\Seff'$
with respect to $\delta g_\pm$ by using the propagators for $\chi$. Notice
that the shift symmetry is not manifest in this series expansion.

The linear term in the effective action is given by
\begin{align}
 i \Seffd{1} &=  \sum_{\alpha= \pm} \int \dd^{d+1} x\,
  \delta g_\alpha(x) W^{(1)}_{\delta g_\alpha} (x)
\,, 
\end{align}
where $W^{(1)}_{\delta g_\alpha}$ is given by the expectation value as
\begin{align}
 & W^{(1)}_{\delta g_+} (x) = - W^{(1)}_{\delta g_-} (x) =  \left\langle
 \frac{\delta i  S_\chi}{\delta g(x)}  \bigg|_{\delta g=0}  \right\rangle\,.
\end{align}
Next, we compute the quadratic terms in $\Seff'$. Taking
the second variation of $\Seff'$ with respect to $\delta g_+$, we obtain
\begin{align}
 W^{(2)}_{\delta g_+ \delta \tilde{g}_+}(x_1,\,x_2) & = i^2 \left\langle
 \frac{\delta S_\chi[\delta g_+,\, \chi_+]}{\delta g_+(x_1)}
 \bigg|_{\delta g_+=0} \frac{\delta S_\chi[\delta g_+,\, \chi_+]}{\delta
 \tilde{g}_+(x_2)} \bigg|_{\delta g_+=0}  \right\rangle_{\pm} \cr
 & \qquad \qquad \qquad \quad  + i \delta(x_1-x_2) \left\langle
 \frac{\delta^2 S_\chi[\zeta_+,\, \chi_+]}{\delta g_+(x_1) \delta \tilde{g}_+(x_1)}
 \bigg|_{\delta g_+=0} \right\rangle_{\pm}\,, \label{Exp:dS++}
\end{align}
where $\delta g$ and $\delta \tilde{g}$ are either $\delta N$, $N_i$, or
$\zeta$, and they can be different metric perturbations. Here, we introduced the expectation value: 
\begin{align}
 & \langle {\cal O}[\chi_+,\, \chi_-] \rangle_{\pm} \equiv \frac{\int D \chi_+
 \int D \chi_- \,{\cal O}[\chi_+,\, \chi_-] e^{i S_\chi[0,\, \chi_+]- i
 S_\chi[0,\,\chi_-]}}{\int D \chi_+ \int D \chi_- \,e^{i S_\chi[0,\,\chi_+]- i
 S_\chi[0,\,\chi_-]}} \,.
\end{align}
Since the action $S_\chi[\delta g_+,\, \chi_+]$ includes only local terms, the variation of
$S_\chi[\delta g_+,\, \chi_+]$ with respect to $\delta g_+(x_1)$ and
$\delta \tilde{g}_+(x_2)$ yields the delta function $\delta(x_1 -x_2)$
in Eq.~(\ref{Exp:dS++}). Similarly, the second variation of $\Seff'$ with respect to $\delta g_-$ is given by 
\begin{align}
  W^{(2)}_{\delta g_- \delta \tilde{g}_-}(x_1,\,x_2) & = i^2 \left\langle
 \frac{\delta S_\chi[\delta g_-,\, \chi_-]}{\delta g_-(x_1)}
 \bigg|_{\delta g_-=0} \frac{\delta S_\chi[\delta g_-,\, \chi_-]}{\delta
 \tilde{g}_-(x_2)} \bigg|_{\delta g_-=0}  \right\rangle_{\pm} \cr
 & \qquad \qquad \qquad \quad  - i \delta(x_1-x_2) \left\langle
 \frac{\delta^2 S_\chi[\delta g_-,\, \chi_-]}{\delta g_-(x_1) \delta \tilde{g}_-(x_1)}
 \bigg|_{\delta g_-=0} \right\rangle_{\pm}\,. \label{Exp:dS--}
\end{align} 
Taking the derivative with respect to both $\delta g_+$ and 
$\delta g_-$, we obtain
\begin{align}
  W^{(2)}_{\delta g_+ \delta \tilde{g}_-}(x_1,\,x_2) & = - i^2 \left\langle
 \frac{\delta S_\chi[\delta g_+,\, \chi_+]}{\delta g_+(x_1)}
 \bigg|_{\delta g_+=0} \frac{\delta S_\chi[\delta g_-,\, \chi_-]}{\delta
 \tilde{g}_-(x_2)} \bigg|_{\delta g_-=0}  \right\rangle_{\pm} \,, \label{Exp:dS+-}
\end{align}
and
\begin{align}
 W^{(2)}_{\delta g_- \delta \tilde{g}_+}(x_1,\,x_2)
 & = - i^2 \left\langle
 \frac{\delta S_\chi[\delta g_-,\, \chi_-]}{\delta g_-(x_1)}
 \bigg|_{\delta g_-=0} \frac{\delta S_\chi[\delta g_+,\, \chi_+]}{\delta
 \tilde{g}_+(x_2)} \bigg|_{\delta g_+=0}  \right\rangle_{\pm} \,. \label{Exp:dS-+}
\end{align}

These functions $W^{(2)}_{\delta g_{\alpha_1} \delta \tilde{g}_{\alpha_2}}(x_1,\, x_2)$ can be
expanded by the propagators of $\chi$ for $\lambda =0$, i.e., the time-ordered
(Feynman) propagator:
\begin{align}
 &  G_F(x_1,\, x_2) \equiv 
 \frac{\int D \chi_+\, \chi_+(x_1) \chi_+(x_2)\,e^{i
 \Scf[\chi_+]}}{\int D \chi_+\,e^{i \Scf[\chi_+]}}\,, 
\end{align}
the anti-time ordered (Dyson) propagator: 
\begin{align}
 &  G_D(x_1,\, x_2) \equiv 
 \frac{\int D \chi_-\, \chi_-(x_1) \chi_-(x_2)\,e^{- i
 \Scf[\chi_-]}}{\int D \chi_-\,e^{i \Scf[\chi_-]}}\,, 
\end{align}
and the Wightman functions: 
\begin{align}
 & G^+(x_1,\, x_2) \equiv 
 \frac{\int D \chi_+ \int D \chi_- \,\chi_-(x_1) \chi_+(x_2) e^{i \Scf[ \chi_+]- i
 \Scf[\chi_-]}}{\int D \chi_+ \int D \chi_- \,e^{i \Scf[\chi_+]- i
 \Scf[\chi_-]}}  \,, \cr
& G^-(x_1,\, x_2) \equiv 
 \frac{\int D \chi_+ \int D \chi_- \,\chi_+(x_1) \chi_-(x_2) e^{i \Scf[ \chi_+]- i
 \Scf[\chi_-]}}{\int D \chi_+ \int D \chi_- \,e^{i \Scf[\chi_+]- i
 \Scf[\chi_-]}} \,.  
\end{align}
Here, $\Scf[\chi]$ denotes the action given by
\begin{align}
  \Scf[\chi] = \frac{1}{2} \int e^{d\rho} \biggl[ \dot{\chi}^2 -
 e^{-2\rho} (\partial_i \chi)^2 - M^2(\phi) \chi^2  \biggr] \dd t \, \dd^d  \bm{x}\,.
\end{align}
Recall that these propagators are mutually related as 
\begin{align}
 & G^-(x_1,\,x_2) = G^{+\,*}(x_1,\,x_2)\,, \label{Exp:Gpm} \\
 & G_F(x_1,\, x_2) =  \theta(t_1 - t_2) G^+(x_1,\, x_2) +  \theta(t_2-
 t_1) G^-(x_1,\,x_2) \,, \label{Exp:GF} \\
 & G_D(x_1,\, x_2) =  \theta(t_1 - t_2) G^-(x_1,\, x_2) +  \theta(t_2-
 t_1) G^+(x_1,\,x_2)\,, \label{Exp:GD}
\end{align} 
where $\theta$ is the Heaviside function.

\subsection{Propagators for $\chi$}  \label{SSec:modefn}
In this subsection, solving the mode function for the heavy field $\chi$, we compute the propagators introduced in the
previous subsection. At the linear order of $\chi$, the equation
of motion is given by
\begin{align}
 & \ddot{\chi}_k + d \dot{\rho} \dot{\chi}_k + \{ M^2(\phi) + (k e^{-\rho})^2 \} \chi_k
 =0\,, \label{Eq:modechi}
\end{align} 
where $\chi_k$ is the Fourier mode of $\chi$. Changing the variable from
$\chi_k$ to $X_k(t)= e^{\frac{d-1}{2} \rho(t)} \chi_k(t)$ and using the conformal
time $\eta$, the mode equation
(\ref{Eq:modechi}) is recast into
\begin{align}
 & X_k'' + \Omega_k^2(\eta)\, X_k =0\,, \label{Eq:modechit}
\end{align}
where the dash denotes the derivative with respect to the conformal time
$\eta$ and the time dependent frequency $\Omega_k$ is given by
\begin{align}
  \Omega_k^2(\eta) = k^2 + (M e^\rho)^2 - \rho'' - {\rho'}^2\,.
\end{align}
Using $W_k$ which satisfies
\begin{align}
 & W_k^2 = \Omega_k^2 + \frac{3}{4} \left( W_k' \over W_k \right)^2 -
 \frac{1}{2} \frac{W_k''}{W_k}\,,  \label{Eq:Wk}
\end{align}
the mode equation (\ref{Eq:modechit}) can be solved as
\begin{align}
 & X_k(\eta) = \frac{1}{\sqrt{2 W_k}} e^{- i \int^\eta \dd
 \eta' W_k(\eta')}\,.   \label{Eq:WKB}
\end{align}

Using the mode function $\chi_k$, we quantize the non self-interacting
heavy field as follows
\begin{align}
 & \chi(x) = \int \frac{\dd^d \bm{k}}{(2\pi)^{d/2}} e^{i \sbm{k} \cdot
 \sbm{x}} a_{\sbm{k}} \chi_k(\eta) + ({\rm h.c.})\,,  \label{Exp:chif}
\end{align}
where $a_{\sbm{k}}$ denotes the annihilation operator. With this
expansion, the Wightman function $G^+(x_1,\,x_2)$ is given by
\begin{align}
 & G^+(x_1,\, x_2) = \int \frac{\dd^d \bm{k}}{(2\pi)^d} e^{i \sbm{k}
 \cdot (\sbm{x}_1- \sbm{x}_2)} \chi_k(\eta_1) \chi_k^*(\eta_2)\,. \label{Exp:G+}
\end{align}
Once the mode function $\chi_k(\eta)$ is given, using
Eqs.~(\ref{Exp:Gpm})-(\ref{Exp:GD}) and (\ref{Exp:G+}), we can
compute all the propagators which appear in the expansion
(\ref{Exp:dS++}).

\section{Ward-Takahashi identity from the scaling symmetry}  \label{Sec:WT}
As was discussed in Sec.~\ref{SSec:scaling}, the action for the single
field model preserves the invariance under the scale transformation, which
changes $\zeta(t,\, \bm{x})$ to $\zeta(t,\, e^{-s} \bm{x}) -s$. This
symmetry is preserved classically also for multi-field models of
inflation, since it is a part of the spatial diffeomorphism. In a
quantum field theory, it is known that a symmetry leads to a
corresponding Ward-Takahashi (WT) identity. When the scaling symmetry is
also preserved at the quantum level, we obtain the WT
identity. In Sec.~\ref{SSec:WT}, we discuss the WT identity from the
scaling symmetry of the correlators of $\chi$. In single field models of
inflation, it was shown that the WT identity of the scaling symmetry yields the consistency relation
which relates the $(n+1)$-point function of $\zeta$ with one soft
external leg to $n$-point function of
$\zeta$~\cite{Maldacena02, PM04, HHK}. Likewise, in Sec.~\ref{SSec:CR}, we
find that the WT identity derived in Sec.~\ref{SSec:WT} gives the
consistency relation which relates the $(n+1)$-point cross-correlation with $n$
$\chi$s and one soft $\zeta$ to the $n$-point auto-correlation of $\chi$.

\subsection{Ward-Takahashi identity}  \label{SSec:WT}
In single field models, an invariant quantity regarding the scale
transformation was constructed in Refs.~\cite{IRgauge_L, IRgauge} by
using the smeared geodesic normal coordinates, defined in
Eq.~(\ref{Def:xg}). Using $\bm{x}_g\equiv \bm{x}_g(t_f)$, evaluated at
the end of inflation $t=t_f$, we define
\begin{align}
 & {^g\!\chi}(t,\, \bm{x}_g) = \chi(t,\, \bm{x})= \chi(t,\, e^{-
 \gbz} \bm{x}_g)\,,
\end{align}
which is invariant under the scaling symmetry with the constant
parameter $s$. Here, $\gbz \equiv \gbz(t_f)$.

When the scaling symmetry is also preserved for the quantum system, 
the correlation functions of ${^g\!\chi}(t,\, \bm{x}_g)$ should be
invariant under the scale transformation as 
\begin{align}
 & \langle \chi_{\alpha_1}(t_1,\, e^{- \gbz}\bm{x}_{g1}) \cdots \chi_{\alpha_n}(t_n,\, e^{- \gbz}\bm{x}_{gn})
 \rangle_{\pm} \big|_{\delta g} \cr
 &\,\,=  \langle \chi_{\alpha_1}(t_1,\, e^{- \gbz+s} \bm{x}_{g1})
 \cdots \chi_{\alpha_n}(t_n,\, e^{- \gbz+s}\bm{x}_{gn}) \rangle_{\pm}
 \big|_{\delta g_s} \label{WT} 
\end{align}
with $\alpha_i = \pm$. Here, $\delta g_s$ denotes the metric
perturbations after the scale transformation. Under the scale
transformation, $\delta N$ and $N_i$ change as 
\begin{align}
 & \delta N_s(t,\, \bm{x}) = \delta N(t,\, e^{-s}\bm{x}) \,, \label{Exp:transN} \\
 & N_{i, s} (t,\, \bm{x}) = e^{-s} N_i(t, e^{-s} \bm{x})\,,   \label{Exp:transNi}
\end{align}
and $\zeta$ changes as in Eq.~(\ref{Exp:transzeta}), and then $\gbz$
changes to $\gbz_s=\gbz-s$. Equation (\ref{WT}) holds only when the
quantum state also preserves the scaling symmetry. This is the
WT identity for the scaling symmetry. At ${\cal O}(s)$,
setting $\delta g=0$, the WT identity yields
\begin{align}
 & \sum_{i=1}^n \bm{x}_i \cdot \partial_{\sbm{x}_i} \langle
 \chi_{\alpha_1}(x_1) \cdots \chi_{\alpha_n}(x_n) \rangle_{\pm}
 -  \int \dd^{d+1} x  \left\langle
 \chi_{\alpha_1}(x_1) \cdots \chi_{\alpha_n}(x_n) \frac{\delta
 i S_\chi[\delta g_+,\, \chi_+]}{\delta \zeta_+(x)} \bigg|_{\delta g_+=0}
 \right\rangle_{\!\pm}  \nonumber \\
 & \qquad \qquad  +  \int \dd^{d+1} x  \left\langle
 \chi_{\alpha_1}(x_1) \cdots \chi_{\alpha_n}(x_n) \frac{\delta
 i S_\chi[\delta g_-,\, \chi_-]}{\delta \zeta_-(x)} \bigg|_{\delta g_-=0}
 \right\rangle_{\!\pm}=0\,.  \label{WTs1} 
\end{align}
Since the changes of $\delta N$ and $N_i$ under the scale transformation
are linear in $N$ and $N_i$ and their derivatives, they vanish after setting $\delta g$ to 0.

Using the WT identity (\ref{WTs1}) with $x_1= \cdots = x_p \equiv x$
and $x_{p+1}= \cdots = x_n \equiv x'$, we obtain
\begin{align}
 & (\bm{x} \cdot \partial_{\sbm{x}} + \bm{x}' \cdot \partial_{\sbm{x}'})
 \langle \chi_\alpha^p(x) \chi_\alpha^{n-p}(x') \rangle_{\pm} 
 - \int \dd^{d+1} y \left\langle \chi_\alpha^p(x) \chi_\alpha^{n-p}(x') \frac{\delta
 i S_\chi[\delta g_+,\, \chi_+]}{\delta \zeta_+(y)} \bigg|_{\delta g_+=0}
 \right\rangle_{\!\pm}  \nonumber \\
 & \qquad \qquad  + \int \dd^{d+1} y \left\langle
  \chi_\alpha^p(x) \chi_\alpha^{n-p}(x') \frac{\delta
 i S_\chi[\delta g_-,\, \chi_-]}{\delta \zeta_-(y)} \bigg|_{\delta g_-=0}
 \right\rangle_{\!\pm}=0 \,,  \label{Exp:WTpr}
\end{align}
where $\alpha= \pm$. In the next section, using these identities, we show the conservation
of $\zeta$, including the loop corrections of the heavy field.

\subsection{Consistency relation (Soft theorem)} \label{SSec:CR}
In single field models of inflation, it is known that the WT identity
for the scaling symmetry gives the consistency relation. The consistency relation for $\zeta$ is an example of the soft
theorem, which was first shown for the soft graviton scattering by 
Weinberg~\cite{WeinbergST}. Recently, Weinberg's soft theorem was
recaptured by Strominger {\it et al.} and was shown to be equivalent to a
Ward-Takahashi identity in an asymptotically flat
spacetime~\cite{Strominger:2013jfa, He:2014laa, Strominger:2014pwa}.

Here, we show that the WT identity (\ref{WTs1})
also gives a consistency relation in multi-field models. Performing the
Fourier transformation of the WT identity (\ref{WTs1}) evaluated at an equal time $t$ with all
$\alpha_i$s chosen as $+$, we obtain  
\begin{align}
 & \left(\sum_{i=1}^n \bm{k}_i \cdot \partial_{\sbm{k}_i} + nd  \right)
 \langle \chi_+(\bm{k}_1) \cdots \chi_+(\bm{k}_n) \rangle_{\pm}  \nonumber \\
 & \qquad   - i \int \dd^{d+1} y \left\langle  \frac{\delta
 S_{\chi}[\delta g_-,\,
 \chi_-]}{\delta \zeta_-(y)} \bigg|_{\delta g_-=0}  \chi_+(\bm{k}_1)
  \cdots \chi_+(\bm{k}_n) \right\rangle_{\!\pm} \nonumber \\
 & \qquad  + i \int \dd^{d+1} y \left\langle \chi_+(\bm{k}_1)
  \cdots \chi_+(\bm{k}_n)  \frac{\delta S_{\chi}[\delta g_+,\,
 \chi_+]}{\delta \zeta_+(y)} \bigg|_{\delta g_+=0}  \right\rangle_{\! \pm}
 =0\,, \label{Eq:WTtf}
\end{align}
where we abbreviated $t$ in the argument of $\chi$s. The correlator in
the first line is simply the in-in $n$-point function of
$\chi(t,\,\bm{k})$. The correlator in the second line is given by the
product of the Wightman function, the Feynman propagator, and the Dyson
propagator, which appear by contracting
$\chi_{\pm}$ with $\chi_{\mp}$, $\chi_+$ with $\chi_+$, and $\chi_-$ with
$\chi_-$, respectively. The correlator in the third line is given by the
product of the Feynman propagator. The interaction vertices inserted at
any time after $t$ are canceled between the terms in the second and third lines.
This ensures the causality in the closed time path formalism. Taking into account
this cancellation, we can rewrite Eq.~(\ref{Eq:WTtf}) as 
\begin{align}
 & \left[ \sum_{i=2}^n \bm{k}_i \cdot \partial_{\sbm{k}_i} + (n-1)d
 \right] \left\langle \chi\left(- \sum_{j=2}^n \bm{k}_j\right) \chi(\bm{k}_2) \cdots
 \chi(\bm{k}_n) \right\rangle' \cr
 & = - i \int^t \dd t_y \int \dd^d \bm{y} \left\langle \left[
 \chi(\bm{k}_1) \cdots \chi(\bm{k}_n),\,  \frac{\delta S_\chi}{\delta
 \zeta(y)} \bigg|_{\zeta=0} \right] \right\rangle' \,,  \label{Eq:CR} 
\end{align} 
where the correlation function with dash denotes the correlation
function from which $(2\pi)^d$ and the delta function are removed, e.g., 
\begin{align}
 & \langle \chi(\bm{k}_1) \cdots \chi(\bm{k}_n) \rangle \equiv (2\pi)^d
 \delta(\bm{k}_1 + \cdots + \bm{k}_n)  \langle \chi(\bm{k}_1) \cdots
 \chi(\bm{k}_n) \rangle' \,. 
\end{align}
In deriving Eq.~(\ref{Eq:CR}), we used
\begin{align}
 & \sum_{i=1}^n \bm{k}_i \cdot \partial_{\sbm{k}_i} \delta(\bm{k}_1 +
 \cdots \bm{k}_n) \langle \chi(\bm{k}_1) \cdots
 \chi(\bm{k}_n) \rangle' \cr
 &  = \delta(\bm{k}_1 +
 \cdots \bm{k}_n) \left( \sum_{i=2}^n \bm{k}_i \cdot
 \partial_{\sbm{k}_i} -d  \right) \left\langle \chi\left(- \sum_{j=2}^n
 \bm{k}_j\right) \chi(\bm{k}_2) \cdots
 \chi(\bm{k}_n) \right\rangle' \,. 
\end{align}

\begin{figure}[t]
\begin{center}
\begin{tabular}{cc}
\includegraphics[width=15cm]{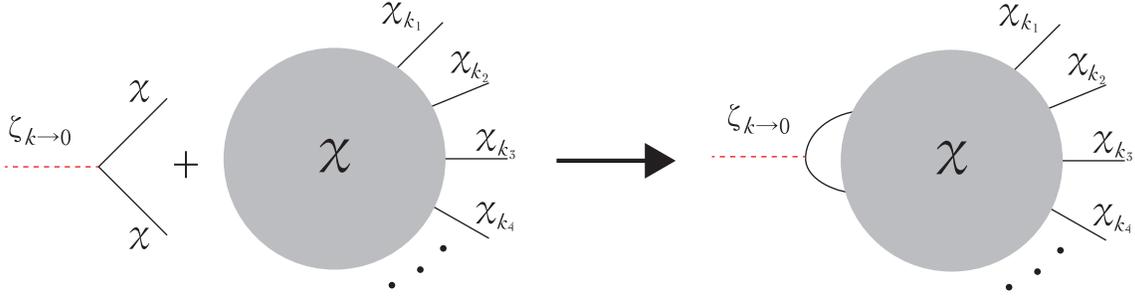}
\end{tabular}
\caption{As an example, we consider a three point interaction vertex in
 $S_{\chi}$ with two $\chi$s and one $\zeta$. Since $\zeta$ is removed
 from the interaction vertex by operating the functional derivative, only two
 $\chi$s remain in the vertex. Contracting these two $\chi$s with $\chi$
 included in the Heisenberg operator $\chi({\bm{k}}_i)$ where $i=1, \cdots n$, 
we obtain the diagram in the right of the arrow. The red dotted line represents the amputated $\zeta$.}   
\label{Fg:CR}
\end{center}
\end{figure} 

The correlation function in the second line of Eq.~(\ref{Eq:CR}) is the
in-in $n$-point function of $\chi(\bm{k})$ where the gravitational
interaction vertices with $n$ heavy fields $\chi$ and one
amputated soft curvature perturbation $\zeta$ are inserted. (See Fig.~\ref{Fg:CR}.) 
Then, attaching the external soft propagator of $\zeta$ to this
correlation function yields the $(n+1)$-point function of $n$ $\chi$s
and one soft $\zeta$, i.e., 
\begin{align}
 &   \left[ \sum_{i=2}^n \bm{k}_i \cdot \partial_{\sbm{k}_i} + (n-1)d
 \right] \left\langle \chi\left(- \sum_{j=2}^n \bm{k}_j\right) \chi(\bm{k}_2) \cdots
 \chi(\bm{k}_n) \right\rangle' \cr
 & \qquad   = - \lim_{k \to 0} \frac{ \langle \zeta(\bm{k}) \chi(\bm{k}_1) \cdots
 \chi(\bm{k}_n) \rangle'}{P_\zeta(k)} \,,
\end{align}
where $P_\zeta(k)$ is the power spectrum of the free $\zeta$. 
This is the consistency relation for the heavy field
$\chi$. The correlation function in the second line contains only one
gravitational interaction vertex with $\zeta$, but it can contain more than one
self interaction vertexes for the heavy field $\chi$. This is one
example of the soft theorem.

Using the WT identity at ${\cal O}(s)$, we derived the consistency
relation for the correlation functions with one soft $\zeta$. Using the
WT identity at ${\cal O}(s^p)$, we can derive the consistency
relations with $p$ soft $\zeta$s.

\subsection{Scaling symmetry and WKB solution}
In this subsection, we explicitly analyze the consistency relation
(\ref{Eq:CR}) for the case with $\lambda=0$. Inserting
Eq.~(\ref{Exp:chif}) into Eq.~(\ref{Eq:CR}), we obtain 
\begin{align}
  &(\bm{k} \cdot \partial_{\sbm{k}} + d) |\chi_k(t)|^2  \cr
  &  = 2 {\rm Im} \biggl[  \int^t
 \dd t' e^{d\rho(t')}   \chi_k^2(t) \left\{
 d(\dot{\chi}_k^*\!^2(t') - M^2 \chi_k^*\!^2 (t') ) - (d-2)
 \frac{k^2}{e^{2\rho(t')}} \chi_k^*\!^2(t') \right\} \biggr].
\end{align} 
Integrating by parts and using the mode equation, we obtain
\begin{align}
 & \bm{k} \cdot \partial_{\sbm{k}} |\chi_k(t)|^2  = 4 \, {\rm Im} \left[
  \int^t \dd t' e^{d \rho(t')} \frac{k^2}{e^{2 \rho(t')}} \chi_k^2(t)
 \chi_k^*\!^2 (t') \right]\,,  \label{CRF}
\end{align}
where $d|\chi_k(t)|^2$ in the left hand side was canceled with the term
which appears by operating the time derivative on the Heaviside function
$\theta(t- t')$.

We can show that the WKB solution satisfies the WT identity
(\ref{CRF}). In order to show this statement, we rewrite Eq.~(\ref{CRF})
as  
\begin{align}
 & \chi_k(\eta) L_k^*(\eta) + \chi_k^*(\eta) L_k(\eta) =0 \,,  \label{CFT2}
\end{align}
introducing 
\begin{align}
 & L_k(\eta) \equiv k \partial_k \chi_k(\eta) - 2 i \chi_k^*(\eta) \int^\eta \dd \eta'
 e^{(d-1)\rho(\eta')} k^2  \chi_k^2(\eta') \cr
 & \qquad \qquad \quad + 2 i \chi_k(\eta) \int^\eta_{\bar{\eta}} \dd \eta'
 e^{(d-1)\rho(\eta')} k^2 |\chi_k(\eta')|^2 + i k \bar{\eta} \chi_k(\eta)\,. \label{Def:Lkt}
\end{align}
The last two terms in $L_k(\eta)$ are canceled between the two terms in
Eq.~(\ref{CFT2}). The time integral of the second term converges by rotating the 
time path as $\eta \to - \infty(1+i\epsilon)$ where $\epsilon$ is a
positive constant. For $L^*_k(\eta)$, the time integral of the second term
should be rotated as $\eta \to - \infty(1-i\epsilon)$. We
choose $\bar{\eta}$ at a time in the distant past when the mode function
can be well approximated by the leading order WKB solution with 
$W_k = k$. (To be precise, $\bar{\eta}$ differs for a different
wavenumber $k$.) Since $L_k(\eta)$ satisfies the mode equation for 
$\chi_k(\eta)$, i.e., 
\begin{align}
 & {L}''_k + (d-1) \rho' L'_k + \{ k^2 + M^2(\phi) e^{2\rho} \} L_k
 =0\,, 
\end{align}
and the initial conditions $L_k(\bar{\eta})= L'_k(\bar{\eta})=0$, $L_k(\eta)$ vanishes all the
time. Thus, we find that the WKB solution satisfies Eq.~(\ref{CFT2}). 

For the exact de Sitter space, in the limit $k e^{-\rho} \ll M$ and 
$H \ll M$, we can easily check that the WKB solution, given by 
\begin{align}
 & \chi_k (t) \simeq \frac{e^{\frac{d}{2}\rho(t)}}{\sqrt{2 M}} \left[ 1
 + \frac{1}{4} \left( \frac{k}{M e^\rho} \right)^2 \left( -1 + i
 \frac{M}{H} \right) \right] e^{- i Mt}\,,
\end{align}
satisfies Eq.~(\ref{CRF}) as
\begin{align}
 & \bm{k} \cdot \partial_{\sbm{k}} |\chi_k(t)|^2 = - e^{(d-1)\rho(t)}
 \frac{k^2}{2 \omega_k^3(t)} \simeq - \frac{e^{-d \rho(t)}}{2 M} \left(
 \frac{k}{M e^\rho} \right)^2 \left\{ 1 - \frac{3}{2} \left(
 \frac{k}{M e^\rho} \right)^2 \right\} \,. 
\end{align}

\section{Conservation of $\zeta$ with loop corrections of heavy field}  \label{Sec:Conservation}
In this section, we show that when the scaling symmetry is preserved,
the curvature perturbation $\zeta$ is conserved in time at super Hubble
scales, including the loop correction of the heavy field $\chi$. For
this purpose, first we rewrite the effective action, using the WT
identities derived in the previous section. Then, using the obtained
effective action, we show the conservation of $\zeta$.

\subsection{Effective action with scaling symmetry}
As discussed for the single field inflation in Sec.~\ref{SSec:single},
the presence of the constant solution is implied by the scaling
symmetry. In this subsection, using the WT identity, we rewrite the
effective action in such a way that the scaling symmetry becomes manifest.

Taking the variation of $S_\chi$ with respect to $\delta g$, we can
compute $W^{(n)}_{\delta g_{\alpha_1} \cdots \delta g_{\alpha_n}}$ and
the effective action. For instance, taking the $n$-th derivative of $S_\chi$
with respect to $\zeta$ and setting $\delta g$ to 0, we obtain  
\begin{align}
 &\frac{\delta^n S_\chi[\delta g,\, \chi]}{\delta
 \zeta(x_1) \cdots \delta \zeta(x_n)} \bigg|_{\delta g=0} \cr
 & = \delta(x_1 - x_2) \cdots \delta(x_{n-1} - x_n) \frac{e^{d \rho(t_1)}}{2}
  \cr  & \quad \times \left[ d^n \! \left( \dot{\chi}^2(x_1) - M^2
 \chi^2(x_1)- \frac{\lambda}{12} \chi^4(x_1)  \right) - (d-2)^n
 e^{-2\rho(t_1)} (\partial_{\sbm{x}_1} \chi(x_1))^2
 \right]. \label{Exp:nS} 
\end{align} 
In this subsection, we show that when the WT identity for $\chi$, given in
Eq.~(\ref{Exp:WTpr}), is fulfilled, the effective action for $\zeta$
preserves the scaling symmetry.

To show this, we further rewrite the WT identity
(\ref{Exp:WTpr}). Operating 
$$
\int \dd^d \bm{x} \int \dd^{d+1} y\, \delta g_\alpha(x) \delta(\bm{x} - \bm{y}) \delta(t_x-t_y)
$$ 
and performing the integration by parts, we obtain
\begin{align}
 & \int \dd^d \bm{x} \langle \chi_\alpha^n (x) \rangle
 \partial_{\sbm{x}} \{ \bm{x} \delta g_\alpha (x) \} + \int \dd^d \bm{x}
 \int \dd^{d+1}y\, \delta g_\alpha(x) \left\langle \chi_\alpha^n(x) \frac{\delta
 i S_\chi}{\delta \zeta_+(y)} \bigg|_{\delta g_+=0} \right\rangle \cr
 & \qquad \qquad - \int \dd^d \bm{x} \int \dd^{d+1}y\, \delta g_\alpha(x) \left\langle
 \chi_\alpha^n(x) \frac{\delta  i S_\chi}{\delta \zeta_-(y)} \bigg|_{\delta g_-=0}
 \right\rangle   = 0\,. \label{Eq:WTnd}
\end{align}
Here, after rewriting 
$\delta(\bm{x}- \bm{y})(\bm{x} \cdot \partial_{\sbm{x}} + \bm{y} \cdot
\partial_{\sbm{y}})$ as 
$\delta(\bm{x}- \bm{y})(\bm{y} \cdot \partial_{\sbm{x}} + \bm{x} \cdot
\partial_{\sbm{y}})$, 
we performed the integration by parts and then we used
$$
 (\bm{y} \cdot \partial_{\sbm{x}} + \bm{x} \cdot \partial_{\sbm{y}})
 \delta(\bm{x}- \bm{y}) =  ( \partial_{\sbm{x}}\, \bm{x} - \bm{x} \cdot \partial_{\sbm{x}})
 \delta(\bm{x}- \bm{y}) = d  \delta(\bm{x}- \bm{y})\,.
$$
Similarly, operating 
$$
\int \dd^d \bm{x} \int \dd^{d+1} y \delta g_\alpha(x) \delta(\bm{x} - \bm{y})
\delta(t_x-t_y) \partial_{x^\mu} \partial_{y^\nu} \,,
$$
on Eq.~(\ref{Def:tWn}) with $n=2$ and $p=1$, where $\mu, \nu=0,\,i$, we obtain  
\begin{align}
 & \int \dd^d \bm{x} \langle \dot{\chi}^2_\alpha (x) \rangle
 \partial_{\sbm{x}} \{ \bm{x} \delta g_\alpha (x) \} + \int \dd^d \bm{x}
 \int \dd^{d+1}y\, \delta g_\alpha(x) \left\langle \dot{\chi}_\alpha^2(x) \frac{\delta
 i S_\chi}{\delta \zeta_+(y)}  \bigg|_{\delta g_+=0} \right\rangle \cr
 & \qquad \qquad - \int \dd^d \bm{x} \int \dd^{d+1}y\, \delta g_\alpha(x) \left\langle
  \dot{\chi}_\alpha^2(x)  \frac{\delta  i S_\chi}{\delta \zeta_-(y)}  \bigg|_{\delta g_-=0}
 \right\rangle   = 0\,, \label{Eq:WTtd} \\
 &  \int \dd^d \bm{x}
 \int \dd^{d+1}y\, \delta g_\alpha(x) \left\langle \dot{\chi}_\alpha(x)
 \partial_i \chi_{\alpha}(x) \frac{\delta
 i S_\chi}{\delta \zeta_+(y)}  \bigg|_{\delta g_+=0} \right\rangle \cr
 & \qquad \qquad - \int \dd^d \bm{x} \int \dd^{d+1}y\, \delta g_\alpha(x) \left\langle
  \dot{\chi}_\alpha(x) \partial_i \chi_\alpha(x)   \frac{\delta  i
 S_\chi}{\delta \zeta_-(y)}  \bigg|_{\delta g_-=0}
 \right\rangle   = 0\,,
\end{align}
and
\begin{align}
 & \int \dd^d \bm{x} \langle \partial_i \chi_\alpha (x)  \partial^i \chi_\alpha (x) \rangle
 \{ \bm{x} \cdot  \partial_{\sbm{x}} + (d-2) \} \delta g_\alpha (x) \cr
 & \qquad \qquad  + \int \dd^d \bm{x}
 \int \dd^{d+1}y\, \delta g_\alpha(x) \left\langle \partial_i \chi_\alpha (x)  \partial^i \chi_\alpha (x)  \frac{\delta
 i S_\chi}{\delta \zeta_+(y)}  \bigg|_{\delta g_+=0} \right\rangle \cr
 & \qquad \qquad - \int \dd^d \bm{x} \int \dd^{d+1}y\, \delta g_\alpha(x) \left\langle
   \partial_i \chi_\alpha (x) \partial^i \chi_\alpha (x)   \frac{\delta
 i S_\chi}{\delta \zeta_-(y)}  \bigg|_{\delta g_-=0}
 \right\rangle   = 0\,, \label{Eq:WTsd}
\end{align}
where we used $\langle \dot{\chi}(x) \partial_i \chi(x) \rangle=0$.
Recalling the expressions of $W^{(1)}_{\delta g_\alpha}$ and
$W^{(2)}_{\delta g_{\alpha_1} \delta g_{\alpha_2}}$, given in
Sec.~\ref{SSec:NaiveEA} and adding Eqs.~(\ref{Eq:WTnd}), (\ref{Eq:WTtd}),
and (\ref{Eq:WTsd}) in such a way that their first terms give $W^{(1)}_{\delta g_\alpha}$, we
obtain 
\begin{align}
 & \int \dd^{d+1} x\, \{\bm{x} \cdot \partial_{\sbm{x}} \delta g_{\pm} (x) \}
 W^{(1)}_{\delta g_{\pm}} (x) \cr
 & \qquad  + \int \dd^{d+1} x \int \dd^{d+1} y\, \delta g_{\pm}(x)
 \left\{ W^{(2)}_{\delta g_{\pm} \zeta_{\pm}}(x,\, y) + W^{(2)}_{\delta
 g_{\pm} \zeta_{\mp}}(x,\, y) \right\} =0 \,.  \label{Eq:WTEA}
\end{align}
As is clear from the derivation, Eq.~(\ref{Eq:WTEA}) also holds, even if
we replace $\delta g_{\pm}(x)$ included in each term with an arbitrary function. Therefore,
replacing $\delta g_{\pm}(x)$ with a constant nonzero number, we obtain
\begin{align}
 & \int \dd^{d+1} x \int \dd^{d+1} y \left\{ W^{(2)}_{\delta g_{\pm} \zeta_{\pm}}(x,\, y) +
 W^{(2)}_{\delta g_{\pm} \zeta_{\mp}}(x,\, y) \right\} = 0 \,.   \label{Eq:WTEAct}
\end{align}

By adding the left hand side of Eq.~(\ref{Eq:WTEA}) multiplied by a constant parameter $- s$ and
Eq.~(\ref{Eq:WTEAct}) with $\delta g_\pm=\zeta_\pm$ multiplied by $- s^2/2$, the linear and the
quadratic terms in the effective action can be given by 
\begin{align}
 &  i \Seffd{1}[\delta g_+,\, \delta g_-]  +  i \Seffd{2}[\delta g_+,\,
 \delta g_-]
 \cr
 & =  \sum_{\alpha = \pm} \int
 \dd^{d+1} x\, \delta g_{\alpha,\, s}(x) W^{(1)}_{\delta g_\alpha} (x) \cr
 & \qquad +  \frac{1}{2!} \!\sum_{\alpha_1, \alpha_2 = \pm}\!  
 \int \dd^{d+1} x_1 \!\int\! \dd^{d+1} x_2 \, \delta g_{\alpha_1,\,
 s}(x_1) \delta \tilde{g}_{\alpha_2,\,s}(x_2) W^{(2)}_{\delta g_{\alpha_1}
 \delta \tilde{g}_{\alpha_2}}(x_1,\,x_2) \cr
 & \qquad + {\cal O}(\delta g^3) \,, \label{Expn:Seffdd}
\end{align}
where $\delta g_s$ are related to $\delta g$ as given in
Eqs.~(\ref{Exp:transzeta}), (\ref{Exp:transN}), and (\ref{Exp:transNi}).
Here, each $\delta g_{i, \alpha}~(i=1,\, 2)$ sums over 
$\delta N_{\alpha, s}$, $N_{i, \alpha, s}$, and $\zeta_{\alpha, s}$.

We can drop the term with one shift vector, because
$W^{(1)}_{N_{i, \alpha}}$, which is proportional to  $\langle \dot{\chi}
\partial_i \chi \rangle$, vanishes. In deriving Eq.~(\ref{Expn:Seffdd}),
we used 
\begin{align}
 & W^{(2)}_{\delta g_{\alpha_1} \delta \tilde{g}_{\alpha_2}}(x_1,\, x_2) =
 W^{(2)}_{\delta \tilde{g}_{\alpha_2} \delta g_{\alpha_1}}(x_2,\,
 x_1)\,,
\end{align}
and $\gbz_+=\gbz_-$, which holds since 
$\zeta_+(t_f,\,\bm{x})=\zeta_-(t_f,\, \bm{x})$. The first term in
Eq.~(\ref{Eq:WTEA}) changes the argument of the metric perturbations in
the linear term of Eq.~(\ref{Expn:Seffdd}). We also changed the arguments of the
quadratic terms, since the modification appears only in higher orders of
$\delta g$.

Equation (\ref{Expn:Seffdd}) shows that with the use of the WT
identity, $\delta g_\alpha(x)$ in $\Seff'$ can be
replaced with $\delta g_{\alpha,\,s}(x)$. Since the rest of the
effective action, $S_{ad}$, is simply
the classical action for the single field model, it also should be invariant
under this replacement. Therefore, when the WT identity (\ref{WT})
holds, the total effective action $S_{\rm eff}$ preserves the invariance
under the change of $\delta g_\alpha$ to $\delta g_{\alpha, s}$.

The effective action (\ref{Expn:Seffdd}) includes the lapse function
and the shift vector. By solving the Hamiltonian and momentum
constraint equations, we can express $\delta N_s$ and $N_{i,s}$ in terms of $\zeta_s$. Using these
expressions, we can eliminate $\delta N_s$ and $N_{i,s}$ in the
effective action as in the single field model~\cite{Maldacena02}. Since the constraint equations for 
$\delta g_s$ are given by replacing $\delta g$ with $\delta g_s$ in the
constraint equations for $\delta g$, the effective action for $\zeta_s$
obtained after eliminating $\delta N_s$ and $N_{i,s}$ should be given
simply by replacing $\zeta$ with $\zeta_s$ in the effective action
expressed only in terms of $\zeta$.

\subsection{Conservation of $\zeta$}

\subsubsection{Tadpole contribution}
Before we discuss the conservation, we show that the linear terms
in the effective action $\Seff$, which is the tadpole terms,
vanish all together. Taking the variation of the effective action with
respect to $N$ and $N_i$, we obtain the constraint equations. The
Hamiltonian constraint for the FRW background gives
\begin{align}
 & d(d-1) \dot{\rho}^2 = \dot{\phi}^2 + \langle \dot{\chi}^2 \rangle +
   e^{-2\rho} \langle (\partial_{\sbm{x}} \chi)^2 \rangle + 2 \langle
 V(\phi,\, \chi) \rangle  \,, 
\end{align}
and at the liner order
\begin{align}
 & (d-1) e^{-2 \rho} \dot{\rho}  \partial^i N_i + \delta N (2
 \langle V \rangle + e^{-2\rho} \langle (\partial \chi)^2 \rangle )-
 d(d-1) \dot{\rho} \dot{\zeta} - \zeta e^{-2 \rho} \langle (\partial
 \chi)^2 \rangle = 0  \label{HconstL}\,,
\end{align}
where $\partial^i \equiv \delta^{ij} \partial_j$. Here, we neglected the
sub-leading contribution at large scales. The scalar part of the momentum constraint gives
\begin{align}
 & \partial_i (\dot{\rho} \delta N - \dot{\zeta})-
 \frac{e^{-2\rho}}{d-1} N_i \langle (\partial_i\chi)^2 \rangle =0\,, \label{MconstL}
\end{align}
where we used $\langle \partial_i \chi \partial_j \chi \rangle \propto \delta_{ij}$. 
The momentum constraint equation can be solved as
\begin{align}
 & \delta N = \frac{\dot{\zeta}}{\dot{\rho}} +
 \frac{e^{-2\rho}}{(d-1)\dot{\rho}} \sum_{i=1}^3 \langle (\partial_i\chi)^2 \rangle
 \partial^{-2} \partial^i N_i \,.  \label{Exp:deltaNwC}
\end{align} 
We can add a homogeneous solution of the Laplace equation on the right hand side.

The action $S_{ad}$ which is accurate at the linear order of $\delta g$
is given by
\begin{align}
  S_{ad} &\simeq  \frac{1}{2} \int \dd^{d+1} x N e^{d(\rho+\zeta)}
 \Bigl[- 2 W(\phi) + \frac{1}{N^2} \bigl\{ - d(d-1) \dot{\rho}^2 +
 \dot{\phi}^2 \bigr\} \cr
 & \qquad \qquad \qquad \qquad \qquad \qquad  -2(d-1) \dot{\rho} \delta N \bigl\{ d  \dot{\zeta}
 - e^{-2(\rho + \zeta)} \partial^i N_i\bigr\}   \Bigr] \,,
\end{align}
where, for our purpose, we partially kept the higher order terms in the
exponential form. The $n$-th order effective action $\Seffd{n}$ includes the local terms
given by 
$$
 \frac{1}{n!} \int \dd^{d+1} x\, \delta g_{1,\, \alpha}(x) \cdots \delta g_{n,\, \alpha}(x)  \left\langle
 \frac{\delta^n i S_\chi[\delta g_\alpha,\, \chi_\alpha]}{\delta
 g_{1,\alpha}(x) \cdots \delta g_{n,\alpha}} \right\rangle  
$$
with $\alpha= \pm$. Adding up these local terms for all $n$, we obtain 
\begin{align}
 & S_{{\rm eff}, {\rm local}}'[\delta g_+,\, \delta g_-]  =
 \langle S_{\chi}[\delta g_+,\, \chi_+] \rangle - \langle
 S_{\chi}[\delta g_-,\, \chi_-] \rangle \,,  
\end{align}
where the terms which do not depend on $\delta g$ are canceled between the
two terms on the right hand side. Adding these local terms to $S_{ad}$
and using the Hamiltonian constraint, we obtain a concise expression as
\begin{align}
 & S_{ad}[\zeta_\alpha]  + \langle S_{\chi}[\zeta_\alpha,\,
 \chi_\alpha] \rangle \cr
 & \simeq - \int \dd^{d+1} x\, N e^{d(\rho + \zeta_\alpha)}
   \left[ 2 \langle V(\phi,\, \chi) \rangle + e^{-2(\rho + \zeta_\alpha)}
 \langle (\partial \chi)^2 \rangle   \right] 
. \label{Exp:SzetaSeff}
\end{align}

As in single field cases, solving the Hamiltonian and momentum
constraint equations, we can express the lapse function and the shift
vector in terms of $\zeta$. Inserting Eq.~(\ref{Exp:deltaNwC}) into the Hamiltonian constraint
(\ref{HconstL}), we find that the degree of freedom to choose the
solution of the Laplace equation changes the
relation between $N_i$ and $\zeta$. Since the domain of integration
extends to the spatial infinity, to regularize the spatial integral, the
perturbed variables should approach to 0 in the spatial
infinity. Therefore, we determine the constant degree of freedom in
Eq.~(\ref{Exp:deltaNwC}), requesting the boundary condition:
\begin{align}
 & \int \dd^d \bm{x} \, \partial^i N_i =0 \,. \label{BCNi}
\end{align} 
With this choice, the Hamiltonian constraint reads 
\begin{align}
 & \int \dd^d \bm{x}\, \delta N \left\{ 2 \langle V \rangle + e^{-2\rho}
 \langle (\partial \chi)^2 \rangle \right\} = \int \dd^d \bm{x} \left\{
 d(d-1) \dot{\rho} \dot{\zeta} + \zeta e^{-2 \rho} \langle (\partial
 \chi)^2 \rangle \right\}\,. 
\end{align}
Using this relation, we can rewrite the action which is valid up to the
linear order as 
\begin{align}
 & S_{ad}[\zeta_\alpha]  + \langle S_{\chi}[\zeta_\alpha,\,
 \chi_\alpha] \rangle \cr
 & \simeq - (d-1) \int \dd^{d+1} x \partial_t \left( \dot{\rho}
 e^{d(\rho+ \zeta_\alpha)} \right) + \int \dd^{d+1} x e^{d(\rho+ \zeta_\alpha)} \! \left\{ (d-1) \ddot{\rho}
 + \dot{\phi}^2 + \langle \dot{\chi}^2 \rangle + \frac{e^{-2\rho}}{d}
 \langle (\partial \chi)^2 \rangle  \right\} \cr
 & \qquad - \int \dd^{d+1} x e^{(d-2) \rho} \left\{ e^{(d-2)\zeta_\alpha} - \left(1
 - \frac{1}{d}\right) e^{d \zeta_\alpha} + \zeta_\alpha \right\} \langle
 (\partial \chi)^2 \rangle \,. \label{Exp:SzetaSeff2}
\end{align}
The first term vanishes as a total derivative. The second term is
proportional to 
\begin{align}
 & (d-1) \ddot{\rho} = - \dot{\phi}^2 - \langle \dot{\chi}^2 \rangle - \frac{1}{d}
 e^{-2\rho} \langle (\partial \chi)^2 \rangle\,,
\end{align}
which can be verified by using the time derivative of the Friedman
equation and the field equations for $\phi$ and $\chi$, given by
\begin{align}
 & \ddot{\phi} + d \dot{\rho} \dot{\phi}+ V_{ph}'(\phi) + \langle \chi^2 \rangle M
 M_\phi =0\,, \label{Eq:H} \\
 & \ddot{\chi}_k + d \dot{\rho} \dot{\chi}_k + \left( M^2 + \frac{\lambda}{2}
 \langle \chi^2 \rangle + k^2 e^{- 2 \rho} \right) \chi_k =0 \,.  \label{Eq:dH} 
\end{align}
The tadpole terms contained in the last line cancel with each other and the
term which does not depend on $\zeta$ is canceled between the action for
$+$ and the one for $-$. In this way, using the background equations and
also choosing the boundary condition for $N_i$ as in Eq.~(\ref{BCNi}), we
can show that the tadpole contributions all disappear.

As we discussed in the previous subsection, the effective action 
$S_{\rm eff}$ stays invariant under the replacement of $\zeta_\alpha(x)$
with $\zeta_{\alpha,s}(x)$. Therefore, the tadpole contribution for
$\zeta_s$ should be given simply by replacing $\zeta(x)$ with
$\zeta_s(x)$ in Eq.~(\ref{Exp:SzetaSeff}). When the background equations are satisfied and $N_{i,s}$ is
chosen to vanish at the spatial infinity (when $N_i$ satisfies the
boundary condition (\ref{BCNi}), $N_{i,s}$ also satisfies it), the terms in the second line of
Eq.~(\ref{Expn:Seffdd}), which are linear in the metric perturbations,
all vanish.

\subsubsection{Existence of constant solution}
Removing the tadpole contribution which vanishes with the use of the
background equations, we only consider the quadratic terms about
$\zeta$. At the linear level, $\zeta_{\alpha, s}$ simply gets the
constant shift as
\begin{align}
 & \zeta_{\alpha,\, s} (x) \simeq \zeta_\alpha(x) -s\,. 
\end{align}
Therefore, the symmetry under the change of $\zeta_\alpha$ into
$\zeta_{\alpha,\, s}$ immediately implies the existence of the constant
solution also in the presence of the loop corrections of the heavy field.

In single filed cases, it is well known that only the constant solution
survives while the other independent solution simply decays in the late
time limit, as far as the background evolution is on an attractor (see,
e.g., Ref.~\cite{JYF}). This fact
explains why the curvature perturbation becomes time independent at
super Hubble scales. When we add a quantum correction from a heavy field, in principle, the ``decaying'' mode can turn
into a growing mode. Such a drastic change of the behaviour of
perturbation can occur, in case the trajectory sizably deviates from the
attractor solution, for instance, owing to an effect of an additional
field. In the present context, the classical background
evolution is determined only by the inflaton and we assume that the
quantum effects of the heavy field always remain to be
perturbative. In such cases, the effect of the heavy field does not drive the ``decaying'' mode to grow in time. Then, the
presence of the constant mode implies the conservation of
the curvature perturbation in time as well as in the presence of the
loop corrections of the heavy field.

\section{Renormalization and scaling symmetry}   \label{Renormalization}
As is common in a non-linear quantum field theory, the effective action for
$\zeta$ potentially diverges due to UV corrections. In our case, the bare coefficients of the effective action 
$W^{(n)}_{\delta g_{\alpha_1} \cdots \delta g_{\alpha_n}}$, which are
expressed in terms of the correlators for $\chi$, can diverge. 
The UV divergence should be renormalized by introducing counter
terms. Depending on a way to introduce the counter terms, the scaling
symmetry might be broken. If it were the case, the WT identity would not hold
any more and the renormalized effective action does not preserve the
scaling symmetry.

When the counter terms are introduced in such a way that the scaling
symmetry is preserved, the WT identity holds also after the UV
renormalization. Then, inserting the WT identity into 
the effective action, which can be renormalized following the standard
procedure since the theory (before the gauge fixing) is a local theory,
and repeating the same argument as we did for the unrenormalized
effective action, we can replace $\zeta_\alpha(x)$ into
$\zeta_{\alpha,\,s}(x)$ in the renormalized effective action.

Since only the heavy field is quantized in computing the
effective action, the curvature perturbation $\zeta$ should be dealt
with as a classical external field. We may set the arbitrary constant
parameter $s$ to a $c$-number variable $\gbz$ in order to express the
effective action in terms of the fluctuation in the local region. 
Then, the effective action includes the non-local contribution
$\gbz$. Nonetheless, the renormalization should proceed in the
standard way, because the inserted non-local contribution, which is
schematically in the following form: 
$$
  0= ({\rm WT~identity, which~identically~vanishes~and~is~local}) \times
  \gbz^n \qquad (n=1,\,2)
$$
is fictitious and does not introduce any non-local interactions.

This aspect may be instructive to speculate on the UV renormalization of
an IR regular quantity. Preserving the scaling symmetry is crucial to cancel
out the potentially IR divergent contribution. In Refs.~\cite{IRgauge_L,
IRgauge}, a quantity which preserves the scaling symmetry was proposed
and it contains non-local contributions. Because of that, in Refs.~\cite{Tsamis:1989yu, Miao:2012xc},
it was suggested that the quantity which preserves the scaling symmetry
may not be able to be renormalized in the standard way by introducing
local counter terms. 

In this paper, we presented a handy example where the UV
renormalization of the heavy field can be performed simply by introducing local counter
terms as well as for a quantity which looks to include a non-local
contribution. Here, we only considered the UV renormalization of the heavy
field $\chi$. It will be important to see if the UV renormalization of
the curvature perturbation also can proceed
by introducing local counter terms or not. We leave this issue for a future
study.

\section{Concluding remarks}  \label{Conclusion}
String theory predicts the presence of a bunch of massive excitations
after reduction to four dimensional spacetime, which may encode, for instance, the information
on the structure of the compactification of the extra dimensions. It is
important to explore a possible imprint of such massive modes on the
curvature perturbation. In this paper, we considered an influence of a
heavy scalar field on the curvature perturbation $\zeta$ at the super Hubble scales. When the mass
of the heavy field $\chi$ is of ${\cal O}(H)$, it can give non-local
radiative corrections to the effective action of $\zeta$, which may
provide a distinctive imprint of the heavy field. We showed
that the time evolution of $\zeta$ at the super Hubble scales is not
affected by the loop corrections of the heavy field as far
as the scaling symmetry, which is entailed in a covariant theory at the
classical level, is
preserved. The implies that the constant adiabatic mode exists as well
as in the presence of the loop corrections of the heavy field.

For simplicity, we considered one heavy field with the standard
canonical kinetic term. However, our argument can be extended in a
straightforward manner to a more general model which contains more than
one heavy fields with a non-canonical kinetic term.

Our result indicates that in order to leave an imprint of massive
fields well after the Hubble crossing, we need to break either of the following conditions~\footnote{Here, we also assume that the (spatially averaged)
background universe is the FRW universe. This excludes, say, the solid
inflation case, where the anisotropic pressure does not vanish in the
large scale limit~\cite{Solid}. (See also Ref.~\cite{Elastic}.)}:
\begin{itemize}
 \item The massive fields do not alter the background evolution at the
       classical level. 
 \item The quantum system preserves the scaling symmetry, which yields
       the Ward-Takahashi identity. 
 \item The radiative corrections of the massive fields on the curvature
       perturbation $\zeta$ are perturbatively suppressed.
\end{itemize}
If the last condition does not hold, we need to perform a
non-perturbative analysis to compute the radiative corrections of the
massive fields.

In this paper, using the WT identity (\ref{WT}) for the scaling symmetry
at ${\cal O}(s)$, we showed that the metric perturbation $\delta g(x)$
in the effective action can be replaced with $\delta g_s(x)$, given in
Eqs.~(\ref{Exp:transzeta}), (\ref{Exp:transN}), and (\ref{Exp:transNi}),
keeping up to the quadratic terms. This argument can be extended to
higher orders in $\delta g$. Using the WT identity (\ref{WT}), we
can derive the WT identity which relates $W^{(n)}$ 
to $W^{(n')}$s with $n'<n$. Adding the WT identity for $W^{(m)}$ with
$m\leq n$ (multiplied by some particular constant factors) to the
effective action $S'_{\rm eff}$, we can replace all $\delta g(x)$ with
$\delta g_s(x)$ in $S'_{\rm eff}$ up to the $n$-th order of
perturbation. After removing the lapse function and the shift vector, we
find that the effective action for the curvature perturbation is
invariant under the replacement of $\zeta(x)$ with 
\begin{align}
 & \zeta_s(x)= \zeta(x) - s - s \bm{x} \cdot \partial_{\sbm{x}} \zeta(x) +
\frac{s^2}{2} (\bm{x} \cdot \partial_{\sbm{x}})^2 \zeta(x) + {\cal O}(s^3)\,. \label{Sol:zetas}
\end{align}
This implies that the curvature perturbation includes a solution which
is given by the $s$-dependent terms in Eq.~(\ref{Sol:zetas}), whose
first term is the constant adiabatic mode. In order to keep the terms
which explicitly depend on $\bm{x}$ perturbatively small, we need to
confine the perturbation within a finite spatial region on each time
slicing. For that, we will need to use other residual \gauge degrees of
freedom, which are addressed in Refs.~\cite{IRgauge_L, IRgauge}.

In this paper, we studied a spin 0 scalar field as the heavy
field. It will be interesting to extend the discussion to include a
field with a more general spin~\cite{NJ15}. Our discussion does not rely on the
explicit form of the interaction vertices nor the propagator. Therefore,
we expect that this extension will be feasible. We leave this study for
a future project~\cite{Spin}.

\acknowledgments
Y.~U. would like to thank N.~Arkani-Hamed, M.~Mirbabayi and M.~Simonovi$\acute{\rm c}$
for interesting discussions about their works, which are relevant to
this work. We are grateful to Y.~Misonoh and S.~Saga for their
participations to the early state of this work. This work is supported
by Grant-in-Aid for Scientific Research (B) No. 26287044. T.~T. was also
supported in part by the Ministry of Education, Culture, Sports, Science
and Technology (MEXT) Grant-in-Aid for Scientific Research on Innovative
Areas, “New Developments in Astrophysics Through Multi-Messenger
Observations of Gravitational Wave Sources”, Nos. 24103001 and 24103006,
and by Grant-in-Aid for Scientific Research (A)  No. 15H02087. Y.~U. is supported by 
JSPS Grant-in-Aid for Research Activity Start-up under Contract
No. 26887018 and the National Science Foundation under Grant No. NSF
PHY11-25915. Y.~U. is partially supported by MEC FPA2010-20807-C02-02
and AGAUR 2009-SGR-168.


\begin{thebibliography}{99}
\bibitem{Ade:2015lrj} 
  P.~A.~R.~Ade {\it et al.} [Planck Collaboration],
  arXiv:1502.02114 [astro-ph.CO].


\bibitem{Ade:2015tva}
  P.~A.~R.~Ade {\it et al.} [BICEP2 and Planck Collaborations],
  Phys.\ Rev.\ Lett.\  {\bf 114} (2015) 101301
  [arXiv:1502.00612 [astro-ph.CO]].

\bibitem{Tolley:2009fg} 
  A.~J.~Tolley and M.~Wyman,
  Phys.\ Rev.\ D {\bf 81}, 043502 (2010)
  [arXiv:0910.1853 [hep-th]].


\bibitem{Jackson:2010cw} 
  M.~G.~Jackson and K.~Schalm,
  Phys.\ Rev.\ Lett.\  {\bf 108}, 111301 (2012)
  [arXiv:1007.0185 [hep-th]].


\bibitem{CW09} 
  X.~Chen and Y.~Wang,
  Phys.\ Rev.\ D {\bf 81}, 063511 (2010)
  [arXiv:0909.0496 [astro-ph.CO]].


\bibitem{Green:2013rd} 
  D.~Green, M.~Lewandowski, L.~Senatore, E.~Silverstein and M.~Zaldarriaga,
  JHEP {\bf 1310}, 171 (2013)
  [arXiv:1301.2630 [hep-th]].


\bibitem{NJ15} 
  N.~Arkani-Hamed and J.~Maldacena,
  arXiv:1503.08043 [hep-th].

\bibitem{Noumi:2012vr} 
  T.~Noumi, M.~Yamaguchi and D.~Yokoyama,
  JHEP {\bf 1306}, 051 (2013)
  [arXiv:1211.1624 [hep-th]].


\bibitem{MM15} 
  M.~Mirbabayi and M.~Simonovi$\acute{\rm c}$,
  arXiv:1507.04755 [hep-th].


\bibitem{Achucarro:2010jv} 
  A.~Achucarro, J.~O.~Gong, S.~Hardeman, G.~A.~Palma and S.~P.~Patil,
  Phys.\ Rev.\ D {\bf 84}, 043502 (2011)
  [arXiv:1005.3848 [hep-th]].


\bibitem{Saito12} 
  R.~Saito, M.~Nakashima, Y.~i.~Takamizu and J.~Yokoyama,
  JCAP {\bf 1211}, 036 (2012)
  [arXiv:1206.2164 [astro-ph.CO]].

\bibitem{Saito13} 
  R.~Saito and Y.~i.~Takamizu,
  JCAP {\bf 1306}, 031 (2013)
  [arXiv:1303.3839, arXiv:1303.3839 [astro-ph.CO]].


\bibitem{Renaux-Petel:2015mga} 
  S.~Renaux-Petel and K.~Turzyński,
  arXiv:1510.01281 [astro-ph.CO].



\bibitem{WeinbergAd} 
  S.~Weinberg,
  Phys.\ Rev.\ D {\bf 67}, 123504 (2003)
  [astro-ph/0302326].

  
\bibitem{WMLL} 
  D.~Wands, K.~A.~Malik, D.~H.~Lyth and A.~R.~Liddle,
  Phys.\ Rev.\ D {\bf 62}, 043527 (2000)
  [astro-ph/0003278].
  
\bibitem{MW03} 
  K.~A.~Malik and D.~Wands,
  Class.\ Quant.\ Grav.\  {\bf 21}, L65 (2004)
  [astro-ph/0307055].
  
\bibitem{LMS} 
  D.~H.~Lyth, K.~A.~Malik and M.~Sasaki,
  JCAP {\bf 0505}, 004 (2005)
  [astro-ph/0411220].
  
\bibitem{LV05} 
  D.~Langlois and F.~Vernizzi,
  Phys.\ Rev.\ D {\bf 72}, 103501 (2005)
  [astro-ph/0509078].
  
  
\bibitem{NS} 
  A.~Naruko and M.~Sasaki,
  Class.\ Quant.\ Grav.\  {\bf 28}, 072001 (2011)
  [arXiv:1101.3180 [astro-ph.CO]].
  


\bibitem{Starobinsky:1986fxa} 
  A.~A.~Starobinsky,
  ``Multicomponent de Sitter (Inflationary) Stages and the Generation of Perturbations,''
  JETP Lett.\  {\bf 42}, 152 (1985)
  [Pisma Zh.\ Eksp.\ Teor.\ Fiz.\  {\bf 42}, 124 (1985)].


\bibitem{Salopek:1990jq} 
  D.~S.~Salopek and J.~R.~Bond,
  ``Nonlinear evolution of long wavelength metric fluctuations in inflationary models,''
  Phys.\ Rev.\ D {\bf 42}, 3936 (1990).
  

\bibitem{SS} 
  M.~Sasaki and E.~D.~Stewart,
  ``A General analytic formula for the spectral index of the density perturbations produced during inflation,''
  Prog.\ Theor.\ Phys.\  {\bf 95}, 71 (1996)
  [astro-ph/9507001].

\bibitem{Sasaki:1998ug} 
  M.~Sasaki and T.~Tanaka,
  ``Superhorizon scale dynamics of multiscalar inflation,''
  Prog.\ Theor.\ Phys.\  {\bf 99}, 763 (1998)
  [gr-qc/9801017].




\bibitem{SZ1210} 
  L.~Senatore and M.~Zaldarriaga,
  JHEP {\bf 1309}, 148 (2013)
  [arXiv:1210.6048 [hep-th]].

\bibitem{ABG} 
  V.~Assassi, D.~Baumann and D.~Green,
  JHEP {\bf 1302}, 151 (2013)
  [arXiv:1210.7792 [hep-th]].




\bibitem{IRgauge_L}
  Y.~Urakawa and T.~Tanaka,
  Phys.\ Rev.\  {\bf D82}, 121301 (2010).
  [arXiv:1007.0468 [hep-th]].


\bibitem{IRgauge}
  Y.~Urakawa and T.~Tanaka,
  Prog.\ Theor.\ Phys.\  {\bf 125}, 1067 (2011)
  [arXiv:1009.2947 [hep-th]].


\bibitem{BGHNT10}
  C.~T.~Byrnes, M.~Gerstenlauer, A.~Hebecker, S.~Nurmi and G.~Tasinato,
  JCAP {\bf 1008}, 006 (2010)
  [arXiv:1005.3307 [hep-th]].

\bibitem{GHT11}
  M.~Gerstenlauer, A.~Hebecker and G.~Tasinato,
  JCAP {\bf 1106}, 021 (2011)
  [arXiv:1102.0560 [astro-ph.CO]].



\bibitem{GS10}
  S.~B.~Giddings and M.~S.~Sloth,
  JCAP {\bf 1101}, 023 (2011)
  [arXiv:1005.1056 [hep-th]].


\bibitem{GS11}
  S.~B.~Giddings and M.~S.~Sloth,
  Phys.\ Rev.\ D {\bf 84}, 063528 (2011)
  [arXiv:1104.0002 [hep-th]].


\bibitem{SZ1203} 
  L.~Senatore and M.~Zaldarriaga,
  JHEP {\bf 1301}, 109 (2013)
  [arXiv:1203.6354 [hep-th]].


\bibitem{PSZ} 
  G.~L.~Pimentel, L.~Senatore and M.~Zaldarriaga,
  JHEP {\bf 1207}, 166 (2012)
  [arXiv:1203.6651 [hep-th]].








\bibitem{IRsingle}
  Y.~Urakawa and T.~Tanaka,
  Prog.\ Theor.\ Phys.\  {\bf 122}, 779 (2009)
  [arXiv:0902.3209 [hep-th]].





\bibitem{SRV1} 
T.~Tanaka and Y.~Urakawa,
  PTEP {\bf 2013}, no. 8, 083E01 (2013)
  [arXiv:1209.1914 [hep-th]].




\bibitem{SRV2} 
  T.~Tanaka and Y.~Urakawa,
  PTEP {\bf 2013}, no. 6, 063E02 (2013)
  [arXiv:1301.3088 [hep-th]].


\bibitem{IRreview} 
 T.~Tanaka and Y.~Urakawa,
  Class.\ Quant.\ Grav.\  {\bf 30}, 233001 (2013)
  [arXiv:1306.4461 [hep-th]].



\bibitem{SRVGW} 
  T.~Tanaka and Y.~Urakawa,
  PTEP {\bf 2014}, no. 7, 073E01 (2014)
  [arXiv:1402.2076 [hep-th]].



\bibitem{Maldacena02} 
  J.~M.~Maldacena,
  JHEP {\bf 0305}, 013 (2003)
  [astro-ph/0210603].
  

\bibitem{PM04} 
  P.~Creminelli and M.~Zaldarriaga,
  JCAP {\bf 0410}, 006 (2004)
  [astro-ph/0407059].



\bibitem{HHK}
  K.~Hinterbichler, L.~Hui and J.~Khoury,
  JCAP {\bf 1401}, 039 (2014)
  [arXiv:1304.5527 [hep-th]].



\bibitem{NFS} 
  M.~H.~Namjoo, H.~Firouzjahi and M.~Sasaki,
  Europhys.\ Lett.\  {\bf 101}, 39001 (2013)
  [arXiv:1210.3692 [astro-ph.CO]].



\bibitem{CNFS} 
  X.~Chen, H.~Firouzjahi, M.~H.~Namjoo and M.~Sasaki,
  Europhys.\ Lett.\  {\bf 102}, 59001 (2013)
  [arXiv:1301.5699 [hep-th]].


\bibitem{Flauger:2013hra} 
  R.~Flauger, D.~Green and R.~A.~Porto,
  JCAP {\bf 1308}, 032 (2013)
  [arXiv:1303.1430 [hep-th]].



\bibitem{GK}
  J.~Ganc and E.~Komatsu,
  JCAP {\bf 1012}, 009 (2010)
  [arXiv:1006.5457 [astro-ph.CO]].

\bibitem{RP} 
  S.~Renaux-Petel,
  JCAP {\bf 1010}, 020 (2010)
  [arXiv:1008.0260 [astro-ph.CO]].


\bibitem{Creminelli:2011rh} 
  P.~Creminelli, G.~D'Amico, M.~Musso and J.~Norena,
  JCAP {\bf 1111}, 038 (2011)
  [arXiv:1106.1462 [astro-ph.CO]].


\bibitem{PJM} 
  P.~Creminelli, J.~Norena and M.~Simonovic,
  JCAP {\bf 1207}, 052 (2012)
  [arXiv:1203.4595 [hep-th]].

\bibitem{ABD12} 
  V.~Assassi, D.~Baumann and D.~Green,
  JCAP {\bf 1211}, 047 (2012)
  [arXiv:1204.4207 [hep-th]].

\bibitem{SZ12} 
  L.~Senatore and M.~Zaldarriaga,
  JCAP {\bf 1208}, 001 (2012)
  [arXiv:1203.6884 [astro-ph.CO]].


\bibitem{Joyce:2014aqa} 
  A.~Joyce, J.~Khoury and M.~Simonovic,
  JCAP {\bf 1501}, no. 01, 012 (2015)
  [arXiv:1409.6318 [hep-th]].



\bibitem{Berezhiani:2013ewa} 
  L.~Berezhiani and J.~Khoury,
  JCAP {\bf 1402}, 003 (2014)
  [arXiv:1309.4461 [hep-th]].
 



\bibitem{Pimentel:2013gza} 
  G.~L.~Pimentel,
  JHEP {\bf 1402}, 124 (2014)
  [arXiv:1309.1793 [hep-th]].


\bibitem{Ghosh:2014kba} 
  A.~Ghosh, N.~Kundu, S.~Raju and S.~P.~Trivedi,
  JHEP {\bf 1407}, 011 (2014)
  [arXiv:1401.1426 [hep-th]].

\bibitem{Kundu:2015xta} 
  N.~Kundu, A.~Shukla and S.~P.~Trivedi,
  arXiv:1507.06017 [hep-th].


\bibitem{Mata:2012bx} 
  I.~Mata, S.~Raju and S.~Trivedi,
  JHEP {\bf 1307}, 015 (2013)
  [arXiv:1211.5482 [hep-th]].

\bibitem{Garriga:2013rpa} 
  J.~Garriga and Y.~Urakawa,
  JCAP {\bf 1307}, 033 (2013)
  [arXiv:1303.5997 [hep-th]].

\bibitem{Garriga:2014ema} 
  J.~Garriga and Y.~Urakawa,
  JHEP {\bf 1406}, 086 (2014)
  [arXiv:1403.5497 [hep-th]].

\bibitem{Garriga:2014fda} 
  J.~Garriga, K.~Skenderis and Y.~Urakawa,
  JCAP {\bf 1501}, no. 01, 028 (2015)
  [arXiv:1410.3290 [hep-th]].



\bibitem{Shiu} 
  K.~Schalm, G.~Shiu and T.~van der Aalst,
  JCAP {\bf 1303}, 005 (2013)
  [arXiv:1211.2157 [hep-th]].


\bibitem{Bzowski:2012ih} 
  A.~Bzowski, P.~McFadden and K.~Skenderis,
  JHEP {\bf 1304}, 047 (2013)
  [arXiv:1211.4550 [hep-th]].


\bibitem{McFadden:2014nta} 
  P.~McFadden,
  JHEP {\bf 1502}, 053 (2015)
  [arXiv:1412.1874 [hep-th]].




\bibitem{IRNG}
  T.~Tanaka and Y.~Urakawa,
  JCAP {\bf 1105}, 014 (2011).
  [arXiv:1103.1251 [astro-ph.CO]].


\bibitem{Creminelli:2011sq} 
  P.~Creminelli, C.~Pitrou and F.~Vernizzi,
  JCAP {\bf 1111}, 025 (2011)
  [arXiv:1109.1822 [astro-ph.CO]].

\bibitem{Pajer:2013ana} 
  E.~Pajer, F.~Schmidt and M.~Zaldarriaga,
  arXiv:1305.0824 [astro-ph.CO].


\bibitem{SeeryLidsey05} 
  D.~Seery and J.~E.~Lidsey,
  JCAP {\bf 0506}, 003 (2005)
  [astro-ph/0503692].
  


\bibitem{Tsamis:1989yu} 
  N.~C.~Tsamis and R.~P.~Woodard,
  Annals Phys.\  {\bf 215}, 96 (1992).

\bibitem{Miao:2012xc} 
  S.~P.~Miao and R.~P.~Woodard,
  JCAP {\bf 1207}, 008 (2012)
  [arXiv:1204.1784 [astro-ph.CO]].

  
\bibitem{Gao:2011mz} 
  X.~Gao,
  JCAP {\bf 1110}, 021 (2011)
  [arXiv:1106.0292 [astro-ph.CO]].

\bibitem{FV} 
  R.~P.~Feynman and F.~L.~Vernon, Jr.,
  Annals Phys.\  {\bf 24}, 118 (1963)
  [Annals Phys.\  {\bf 281}, 547 (2000)].

\bibitem{Wu:2006xp} 
  C.~H.~Wu, K.~W.~Ng, W.~Lee, D.~S.~Lee and Y.~Y.~Charng,
  JCAP {\bf 0702}, 006 (2007)
  [astro-ph/0604292].


\bibitem{KKT} 
  N.~Katirci, A.~Kaya and M.~Tarman,
  JCAP {\bf 1406}, 022 (2014)
  [arXiv:1402.3316 [hep-th]].


\bibitem{FH}
  R.~P.~Feynman and A.~R.~Hibbs, Quantum Mechanics and Path Integrals
	(McGraw-Hill, New York, 1965)

\bibitem{CL1982} 
  A.~O.~Caldeira and A.~J.~Leggett,
  Physica {\bf 121A}, 587 (1983).




\bibitem{WeinbergST} 
  S.~Weinberg,
  Phys.\ Rev.\  {\bf 140}, B516 (1965).

\bibitem{Strominger:2013jfa} 
  A.~Strominger,
  JHEP {\bf 1407}, 152 (2014)
  [arXiv:1312.2229 [hep-th]].

\bibitem{He:2014laa} 
  T.~He, V.~Lysov, P.~Mitra and A.~Strominger,
  JHEP {\bf 1505}, 151 (2015)
  [arXiv:1401.7026 [hep-th]].

\bibitem{Strominger:2014pwa} 
  A.~Strominger and A.~Zhiboedov,
  arXiv:1411.5745 [hep-th].


\bibitem{JYF} 
  J.~Garriga, Y.~Urakawa and F.~Vernizzi,
  arXiv:1509.07339 [hep-th].

\bibitem{Solid} 
  S.~Endlich, A.~Nicolis and J.~Wang,
  JCAP {\bf 1310}, 011 (2013)
  [arXiv:1210.0569 [hep-th]].

\bibitem{Elastic} 
  A.~Gruzinov,
  Phys.\ Rev.\ D {\bf 70}, 063518 (2004)
  [astro-ph/0404548].


\bibitem{Spin}
  Y.~Urakawa {\it et al.},~ in preparation
%


\end{thebibliography}
\end{document}